\documentclass[article,preprint]{elsarticle}
\usepackage{color}

\journal{Journal of \LaTeX\ Templates}

\usepackage{rotating}
\usepackage{setspace}
\usepackage{color}
\usepackage{enumitem}
\usepackage{amsmath}
\usepackage{ulem}
\begin{document}

\begin{frontmatter}

\title{Small-Angle X-ray Scattering Unveils the Internal Structure of Lipid Nanoparticles}

\author[affa]{{Francesco Spinozzi}
\corref{mycorrespondingauthora}}
\cortext[mycorrespondingauthora]{Corresponding author}
\ead{f.spinozzi@univpm.it}
\author[affa]{Paolo Moretti}
\author[affb]{Diego Romano Perinelli}
\author[affc]{Giacomo Corucci}
\author[affa]{Paolo Piergiovanni}
\author[affd]{Heinz Amenitsch}
\author[affe]{Giulio Alfredo Sancini}
\author[afff]{Giancarlo Franzese}
\author[affg]{{Paolo Blasi}
\corref{mycorrespondingauthorb}}
\cortext[mycorrespondingauthorb]{Corresponding author}
\ead{p.blasi@unibo.it}

\address[affa]{Department of Life and Environmental Sciences, Polytechnic University of Marche, Italy}
\address[affb]{School of Pharmacy, University of Camerino, Camerino, Italy}
\address[affc]{{Institut Laue-Langevin, Grenoble, France} and {\'Ecole Doctorale de Physique, Universit\'e Grenoble Alpes, Saint-Martin-d'H\'eres, France} and {Department of Chemistry, Imperial College London, London, UK}}
\address[affd]{Institute for Inorganic Chemistry, Graz University of Technology, Graz, Austria}
\address[affe]{School of Medicine and Surgery, University of Milan Bicocca, Milan, Italy}
\address[afff]{Secci\'o de F\'isica Estad\'istica i Interdisciplin\`aria - Departament de F\'isica de la Mat\`eria Condensada, \& Institut de Nanoci\`encia i Nanotecnologia, Universitat de Barcelona,
            Mart\'i i Franqu\`es 1,
            Barcelona,
            08028,
            Spain}
\address[affg]{Department of Pharmacy and Biotechnology, University of Bologna, Bologna, Italy}

\begin{abstract}
Lipid nanoparticles own a remarkable potential in nanomedicine, only
partially disclosed. While the clinical use of liposomes and cationic
lipid-nucleic acid complexes is well-established, liquid lipid
nanoparticles (nanoemulsions), solid lipid nanoparticles, and
nanostructured lipid carriers have even greater possibilities. However,
they face obstacles in being used in clinics due to a lack of
understanding about the molecular mechanisms controlling their drug
loading and release, interactions with the biological environment
(such as the protein corona), and shelf-life stability. To create
effective drug delivery carriers and successfully translate bench
research to clinical settings, it is crucial to have a thorough
understanding of the internal structure of lipid
nanoparticles. Through synchrotron small-angle X-ray scattering
experiments, we determined the spatial distribution and internal
structure of the nanoparticles' lipid, surfactant, and the bound water in
them. The nanoparticles themselves have a barrel-like shape that
consists of coplanar lipid platelets (specifically cetyl palmitate)
that are covered by loosely spaced polysorbate 80 surfactant molecules,
whose polar heads retain a large amount of bound water.
To reduce the interface cost of bound water with unbound
  water without stacking, the platelets collapse onto each
  other.
This internal structure challenges the
classical core-shell model typically used to describe solid lipid
nanoparticles and could play a significant
role in drug loading and release, biological fluid interaction, and
nanoparticle stability, making our findings
valuable for the rational design of lipid-based nanoparticles.
\end{abstract}


\end{frontmatter}

\section{Introduction}
Lipid nanoparticles (LNPs) have been widely investigated as drug
delivery systems for enhancing drug bioavailability and targeting
therapeutic and diagnostic agents to pathological sites such as brain
and solid tumors~\cite{Muller2000, Tapeinos2017, Xu2022, Tenchov2021,Dalmagro2018,blasi2013b}.
The recent introduction to clinics of RNAi and mRNA-based medicinal
products using LNPs~\cite{Akinc2019, Hou2021} has highlighted the
enormous potential of lipid carriers as drug delivery systems for both
large biomacromolecules like nucleic acid and peptides, as well as
small molecule drugs. However, LNPs comprise a diverse range of
nanometer carriers composed of lipid molecules. Indeed, due to the
broad definition of lipids according to IUPAC~\cite{Moss1995}, LNPs
encompass various structurally different nanoscale carriers, including
liposomes, liquid LNPs, solid LNPs, nanostructured lipid carriers, and
cationic lipid-nucleic acid complexes~\cite{Tenchov2021}.

LNPs for drug delivery have the advantage of using GRAS
materials~\cite{Doktorovova2014} and industrial-scale production
protocols~\cite {Tenchov2021}, which increases the likelihood of
developing effective nanotechnology-based medicine for clinical use.
However, the lack of a deep and comprehensive understanding of the LNP
structure hinders the rational, safe and effective design of these
drug carriers. The effectiveness and safety of LNPs are not only
influenced by the lipids in their formulation and the amount of drug
they can hold but also by various factors such as their size, shape,
surface chemistry, internal structure, and drug distribution. Careful
analysis of the entire system is essential in comprehending the
nano-bio interface, which is accountable for the safety and efficacy
of nanotechnology-based medication.~\cite{Mitchell2021}.

 Solid LNPs are a type of drug delivery and targeting
  carriers that have shown great promise due to their stability over
  time.  Compared to other lipid-based delivery systems
    like liposomes and nanoemulsions, solid LNPs have the solid-state
    stability of the core that is less prone to problems such as drug
    leakage/degradation and particle coalescence. They can encapsulate
    a variety of hydrophobic and hydrophilic
    drugs~\cite{Seo2023,Haddadzadegan2022} and this adaptabity to
    deliver a wide range of therapeutic compounds is expected to
    increase demand for them in the market~\cite{market}.

Although solid LNPs have shown excellent performance in preclinical
studies, they have been studied as nanoscopic carriers for drug
delivery and targeting for only the last three decades, much less than
liposomes and cationic lipid particles for RNA delivery. Also, solid
LNPs still face stability challenges like premature drug leakage and
nanoparticle aggregation, which hinder their clinical
use~\cite{Tenchov2021}. Researchers previously thought lipid
polymorphism was responsible for these issues, as observed through
techniques like calorimetry and X-ray diffraction.  Solid LNPs were
described using a core-shell model with a (solid) lipid core
stabilized by a surfactant shell, possibly penetrating with its
hydrophobic tail the lipid surface~\cite{Campani2018}. However, recent
findings have shown that the interplay between lipids and surfactants
is more complex and LNP structure cannot be
explained by this model alone~\cite{Pink2019, Pink2021}. This new
understanding sheds light on the structure, shelf-life stability, drug
loading/release, and interaction with the biological environment of
solid LNPs, offering new possibilities for drug delivery.

In this study, we used solid LNPs made of cetyl palmitate (CP) and
polysorbate 80 (P80) to investigate how the combination of lipids and
surfactants affects the internal lamellar structure and the P80
surface coverage. CP was selected due to its easy biodegradability in
vivo~\cite{Weyhers2006,Doktorovova2014}, fundamental to avoiding waste
disposable, while P80 is non-ionic surfactant approved by regulatory
agencies for parenteral use and so already employed in injectable
formulations~\cite{Jones2018}.

Understanding the internal structure and composition is crucial for
predicting drug loading, cargo stability, and release based on the
drug's physicochemical properties. Similarly, studying the surface
characteristics is essential for analyzing the nano-bio interface and
comprehending the role of adsorbed biomolecules (bio-corona) on
biodistribution and cellular uptake.

We conducted synchrotron small-angle X-ray scattering (SAXS)
experiments on {{\rm P80}} micelles and {{\rm P80}}-stabilized solid
lipid nanoparticles at different concentrations and temperatures. By
using advanced methods, we were able to determine that these particles
have a barrel-like shape made up of {{\rm CP}} platelets that are
covered by loosely spaced {{\rm P80}}
molecules retaining a large amount of bound water. These
  findings demonstrate the interplay between lipid,
surfactant, and water in the formation of the LNP inner
core. Furthermore, $\approx 65$\% of the platelet surface
  is made of water bound to P80 and in contact with amorphous {{\rm CP}}.
Consequently, we found that some lipid regions
are in contact with the surrounding water
via bound water.

\section{Materials and methods}
\label{mam}
\subsection{Materials}

{{\rm CP}} (batch 120851, purity $\sim 93\%$) was kindly gifted by
Gattefoss\'e s.a.s. (Saint-Priest, France) while {{\rm P80}} (batch
BCBV8843) was from Sigma-Aldrich (Milan, Italy). Water (resistivity
$18.3$~M$\Omega$cm at 25$^\circ$~C) was produced with a
Synergy\textsuperscript{\textregistered} UV Water Purification System
(Millipore Sigma, USA). If not specified, all the materials and
solvents used in the present research work were used as provided by
the supplier without further purification.

\subsubsection{{Solid LNPs preparation}}
\label{lnp}
Solid LNPs were prepared through the hot, high-pressure homogenization
technique with slight adaptations of a previously reported
protocol~\cite{blasi2011,blasi2013}.  Briefly, 4~g of {{\rm CP}},
melted at 65~$^\circ$C, were slowly added to 40~mL of heated water
(65~$^\circ$C) containing {{\rm P80}} at a concentration of 2\% (w/v)
under mixing at 8000~rpm by a high-shear mixer (Ultra Turrax T25
IKA\textsuperscript{\textregistered} Werke GmbH \& Co. KG, Staufen,
Germany). The obtained emulsion was passed through a homogenizer
(high-pressure homogenizer Emulsiflex C5, Avestin Inc., Ottawa,
Canada) 7~times at a pressure of 1500~bar~\cite{blasi2011}. The
homogenizer was conditioned at 65~$^\circ$C during all the
homogenization process. After the last homogenization cycle, the
obtained nanoemulsion was cooled down in an ice bath, maintaining the
dispersion under mild magnetic stirring (20~min). Upon cooling, the
nanoemulsion droplets solidify, generating solid LNPs.

\subsection{Methods}

\subsubsection{DLS experiments}
Dynamic Light Scattering (DLS) experiments were carried out to
evaluate the average size, at micrometric resolution, of solid LNPs as
well as their stability as a function of the time from
preparation. Measurements were performed on a Zetasizer PRO instrument
(Malvern Panalytical Ltd, Malvern, United Kingdom) at 25~$^\circ$C by
detecting the intensity of the light (wavelength 6328~{\AA}) scattered
at a fixed angle of 173$^\circ$. A freshly prepared dispersion of LNPs
was diluted to 1~{g/L}, and three independent DLS measurements of the
second-order intensity autocorrelation functions, $g_2(\tau)-1$, where
$\tau$ is the correlation time, were performed after 0, 2, 6, 15, and
30 days passed from the nanoemulsion preparation. Data were analyzed
by assuming a Gaussian distribution of the hydrodynamic LNP radius,
$R_H$, as detailed in the Sect.~S1 of the
Supplementary Material (SM).  Zeta Potential measurements were also
performed using the same instrument.

\subsubsection{AFM experiments}
AFM measurements were carried out on an AIST-NT Scanning Probe
Microscopy (Horiba Scientific, Kyoto, Japan). Images were generated in
non-contact mode with a pyramidal silicon tip with radius 80~{\AA}. To
improve the quality of the measurements, samples were diluted to
0.1~{g/L}. An amount of $\approx 5$~$\mu$L of the diluted dispersion
was deposited on a freshly cleaved mica surface and then dried with a
nitrogen flux. All images were acquired with a resolution of
$512\times 512$ pixels at a scan rate of 1~Hz and were analyzed with
Gwyddion~\cite{Necas2012} and ImageJ~\cite{Schneider2012} software.
The AFM particle size analysis was carried out by selecting about 50
individual LNPs and measuring the distance $R_{\rm c}$ between the
center and the border along randomly oriented straight lines passing
through the center of the particle. A histogram of all measurements
was then determined by using a 50~{\AA} grid and fitted using a simple
Gaussian distribution.

  \subsubsection{SAXS experiments}

SAXS experiments were carried out at the beamline ID02 of ESRF, the
European Synchrotron Radiation Facility (Grenoble, France). A unique
flow-through capillary, with quartz walls of 10~$\mu$m and a diameter
of $\sim 2.0$~mm, equipped with a motorized syringe that allowed the
sample volume to be moved continuously forward and backward in order
to limit the radiation damage, was used for both samples and
buffers. Two sample-to-detector distances were used, corresponding to
1.5~m and 15~m, and data were merged to achieve a $q$-range ($q=4\pi
\sin\theta/\lambda$ being the modulus of the scattering vector, where
$2\theta$ is the scattering angle and $\lambda=0.995$~{\AA} the X-ray
wavelength) of $0.001-0.5$~{\AA}$^{-1}$. For each of the two
distances, SAXS measurements were performed at the temperature of 20,
25, 30, 37, 25 and 20~$^\circ$C by using an increasing and decreasing
temperature ramp accessible using a Peltier-controlled stage. 2D SAXS
patterns were collected by using a CCD detector (Rayonix MX170 HS) and
subsequently corrected for the CCD dark counts, for the spatial
inhomogeneities of the detector and normalized to an absolute scale
using the standard procedure~\cite{Narayanan2018}. Ten 2D SAXS
patterns of 0.1~s duration were collected for each sample or
buffer. The 1D SAXS profiles were obtained by azimuthally averaging
each of the 10 normalized 2D SAXS patterns. The mean and the standard
deviation of the 1D SAXS profiles were calculated based on the 10 2D
SAXS patterns. To each sample, the buffer contribution, multiplied by
the factor $1-\eta$, $\eta$ being the sample volume fraction, was
subtracted from the 1D SAXS profile to finally obtain the macroscopic
differential scattering cross-section, $d\Sigma/d\Omega(q)$, together
with its standard deviation, $\sigma(q)$, as a function of $q$.

Other SAXS experiments on a second batch of samples prepared with the
same method exposed in the Sect.~2.1.1 were performed at the
Austrian SAXS beamline of the ELETTRA synchrotron (Trieste,
Italy). Measurements of both samples and buffers were carried out in a
unique quartz capillary (diameter 1.5~mm and wall thickness 10~$\mu$m)
mounted on a thermostatic support connected to a circulation bath for
temperature control. 2D SAXS patterns were collected 3~times with an
acquisition time of 20~s using a Pilatus3~1~M detector. Data
reduction was performed with the methodology previously described for
the ESRF data.

\subsubsection{{SAXS models}}
\label{sm}
We have developed novel models to analyze SAXS data of solid lipid
nanoparticles formed by cetyl palmitate and stabilized by
polysorbate~80 (Fig.~1) as well as SAXS data of only
{{\rm P80}}. The models take into account the whole $q$-range of all
synchrotron SAXS data and exploit the information coming from (i) the
absolute calibration of such data, (ii) the chemical compositions of
{{\rm CP}} and {{\rm P80}} (Table~1) and (iii) their
nominal concentrations in the SAXS investigated water
solutions. Moreover, those models are applied to simultaneously fit
all the experimental SAXS curves by following a so-called global fit
approach~\cite{spinozzi2014}.

Data of samples containing only {{\rm P80}} have been analyzed with
the form factor of cylinders with spherical end-caps~\cite{kaya2004},
with size distribution described by the ladder
model~\cite{thomas1997}, and with the structure factor derived by a
perturbation of the Percus-Yevick (PY) model due to the hard sphere
double Yukawa potential (HSDY) in the framework of the random phase
approximation (RPA)~\cite{Hansen1976,PY_closure,spinozzi2020}.

SAXS curves of LNPs have been modeled by the form factor of a barrel
formed by the stacking of polydisperse {{\rm CP}}
platelets~\cite{Schmiele2013,Schmiele2015} covered with a non
continuous layer of {{\rm P80}} with low surface density. The stacking
structure factor has been described in the framework of the
para-crystal theory~\cite{Hosemann1962, Hosemann1952, Hosemann1967,
  Matsuoka1987}. The excess {{\rm P80}} molecules of these samples are
considered to form micelles, described with the same approach used for
samples of only {{\rm P80}}.

\begin{table}[h]
  \begin{center}
\begin{tabular}{lcccccccc}
  &
 ${\rm >C=}$ &
 ${\rm =O}$ &
 ${\rm -O-}$ &
 ${\rm OH}$ &
 ${\rm CH}$ &
 ${\rm CH_2}$ &
  ${\rm CH_3}$&
  ${\rm H_2O}$ \\
  \hline
  n. electrons &
  6 & 8 & 8 & 9 & 7 & 8 & 9 & 10\\
  $\nu^\circ$ $^{(\rm a)} $({\AA}$^3$) &
  13.0 & 12.0 & 15.0 & 16.0 & 21.5 & 27.7 & 52.9 & 29.9\\
  \hline
{{\rm CP}} polar head &
1&1 & 1 &   &   &   &  & \\
{{\rm CP}} hydrophobic tail &
 &  &   &   &   & 29 &2 &\\
{{\rm P80}} dry polar head &
1 &1 & 22 & 3 & 4 & 42 & &\\
{{\rm P80}} hydrophobic tail &
 &  &   &   & 2 & 14 &1 &\\
\end{tabular}
  \end{center}
\caption{Chemical groups forming the polar and the hydrophobic domains
  of {{\rm CP}} and {{\rm P80}} molecules. The first bloc of the table reports
  the number of electrons and the molecular volume at 25~$^\circ$C of
  each group. The second bloc reports the abundance of the groups in
  the hydrophobic and polar domains {{\rm CP}} and {{\rm P80}}
  molecules. $^{(\rm a)}$ Data calculated according to
  Marsh et al.~\cite{Marsh2010}.}
\label{groups}
\end{table}

\begin{figure}
  \begin{center}
  \includegraphics[width=0.35\textwidth]{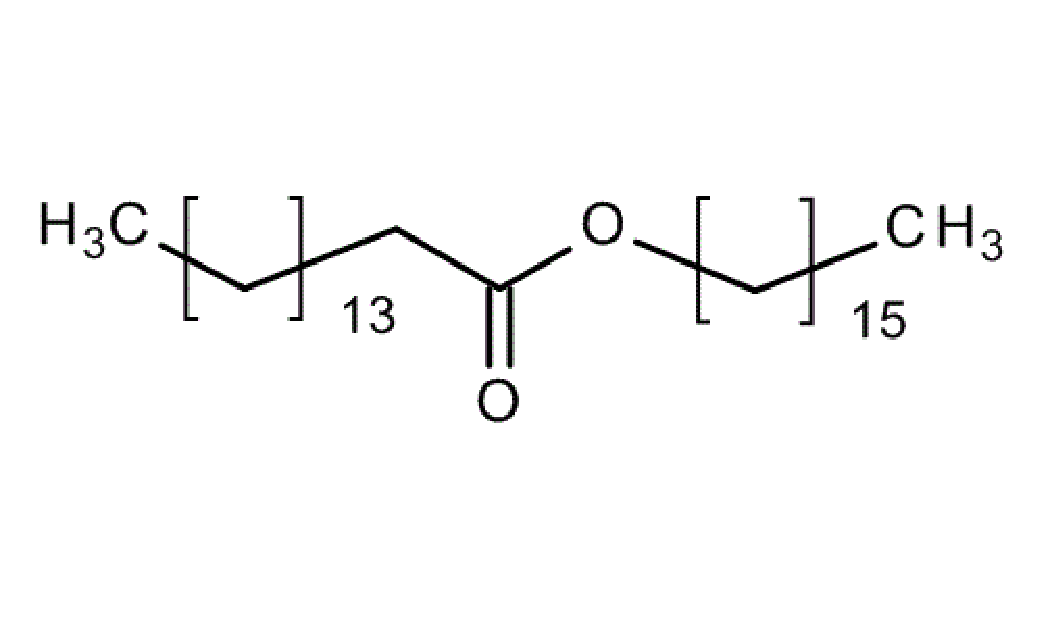}
  \vskip 0.3 cm
  \includegraphics[width=0.60\textwidth]{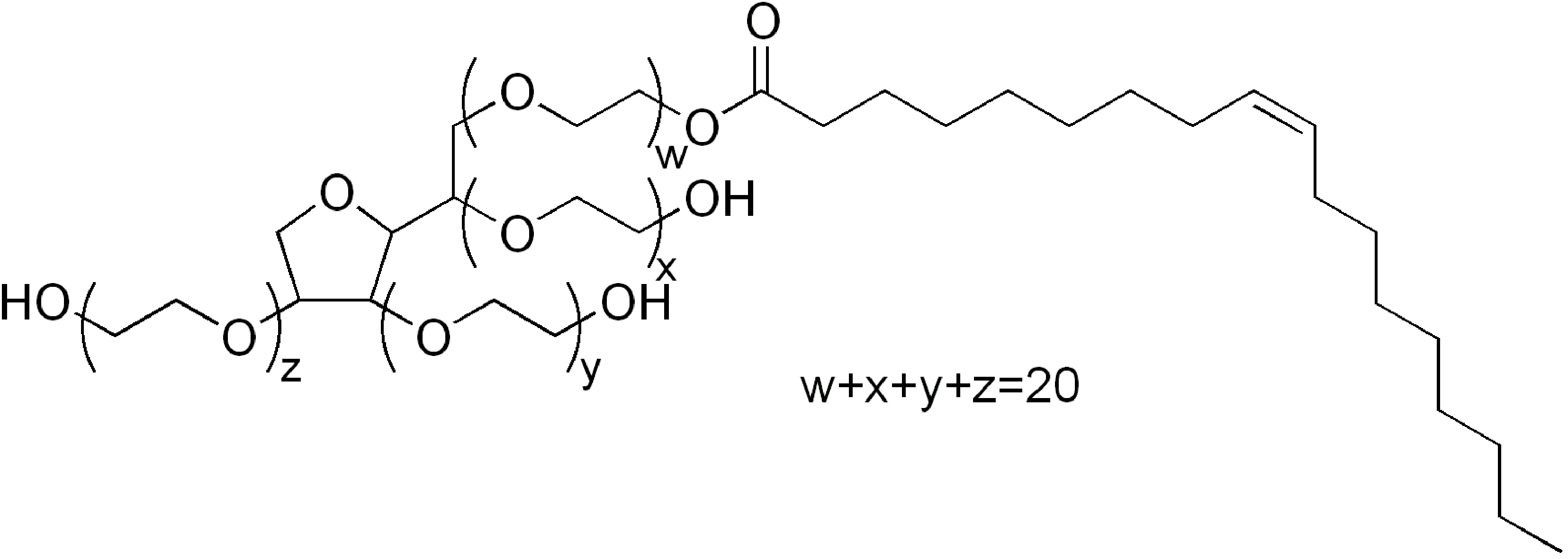}
  \end{center}
  \caption{Chemical structure of the molecules cetyl palmitate,
{{\rm CP}} (top), and polysorbate~80, {{\rm P80}} (bottom).}
  \label{molecules}
\end{figure}

In the following paragraphs, a complete description of these models is shown.

\paragraph{{SAXS of growing and interacting end-capped cylindrical micelles}}
\label{saxs1}
Micelles composed by the nonionic surfactant {{\rm P80}} are supposed
to be distributed in different sizes according to the ladder model
derived by Thomas et al.~\cite{thomas1997}. We first consider the
chemical potential of a micelle formed by $m$ self-assembled
molecules, $\mu_m =\mu^\circ_m+{{\rm R}}T\log C_m$, where ${{\rm R}}$
is the perfect gas constant, $T$ the absolute temperature, $C_m$ the
molar concentration of the micelle and $\mu^\circ_m$ is the standard
chemical potential in the molar unit (corresponding to $C_m=1$~M). The
formation of this micelle from $m$ isolated molecules is written as a
chemical reaction, $ m \, {\rm P80} \rightleftharpoons {\rm P80}_m
$. At the equilibrium, according to standard thermodynamics, the
chemical potential of {{\rm P80}} in any state should be the same,
hence $\mu_1 =\mu_m/m$. It follows that $C_m=C_1^m
e^{-(\mu^\circ_m-m\mu^\circ_1)/({{\rm R}}T)}$. The ladder
model~\cite{thomas1997} simply assumes that the standard chemical
potential difference, $\mu^\circ_m-m\mu^\circ_1$, is a linear function
of $m$,
\begin{eqnarray}
\mu^\circ_m-m\mu^\circ_1 = \Delta+(m-m_0)\delta
\end{eqnarray}
where $\Delta$ is the free energy gain when a micelle with the minimum
aggregation number $m_0$ is formed and $\delta$ is the free energy
gain when a molecule is added to a micelle already formed. To note,
both $\Delta$ and $\delta$ must be negative, indicating that the two
corresponding processes are favored. On the other hand,
$\Delta-m_0\delta$, the free energy required to form two end-caps in
the cylindrical body of the micelle should be positive. The mass
balance of {{\rm P80}} leads to the following equation, $ C_{\rm
  P80}=C_1+\sum_{m=m_0}^\infty m C_m$, where $C_{\rm P80}=c_{\rm
  P80}\,d_{\rm wat}\,/M_{{\rm P80}}$ is the nominal molar
concentration of {{\rm P80}} ($c_{\rm P80}$ is the w/v concentration
at the reference temperature $T_\circ=298.15$~K, $M_{{\rm P80}}$ is
the molecular weight of {{\rm P80}} and $d_{\rm wat}$ is the bulk
water relative mass density, calculated, according to Eq.~2 of
Spinozzi et al.~\cite{spinozzi2020}, as a function of $T$).  We thus
derive $C_{\rm P80}=C_1+e^{-(\Delta-m_0\delta)/({{\rm R}}T)}
\sum_{m=m_0}^\infty m C_1^m e^{-m \delta/({{\rm R}}T)} $. The last
equation can be re-written in terms of the the fraction of free {{\rm
    P80}} molecules in solution, $\alpha_1=C_1 /C_{\rm P80}$, and by
calculating the derivative of the sum of the first $m$ elements of a
geometric series. The result leads to an equation of the unique
variable $z=\alpha_1 C_{\rm P80} e^{-\delta/({{\rm R}}T)}$,
\begin{eqnarray}
  z  e^{\delta/({{\rm R}}T)}  +e^{-(\Delta-m_0\delta)/({{\rm R}}T)}  z^{m_0} \frac{m_0-z(m_0-1)}{(1-z)^2} &=& C_{\rm P80}
  \label{zeta}
\end{eqnarray}
We have checked that, by assuming $z<1$, Eq.~2 can by
numerically solved. As a result, the fraction $\alpha_1$ can be
obtained as a function of $c$, $T$ and the two thermodynamic
parameters ruling the micellar processes, $\Delta$ and
$\delta$. Moreover, the average micellar aggregation number,
$<\!m\!>$, can be easily derived, according to Eq.~S10 of the {SM}.
Examples of numerical solutions of Eq.~2 and calculation of
$C_m$ are shown in Fig.~S1 of the {SM}.  By extending this treatment
from a discrete to a continuous approach and by neglecting the SAXS
contribution of isolated {{\rm P80}} molecules, the average form
factor of end-cap cylindrical ({{\rm ec}}) micelles can be written by
\begin{eqnarray}
  P_{{\rm ec}}(q )&=&\int_{m_0}^\infty p(m)  P_{{\rm ec},m}(q ) dm
\label{PP}
\end{eqnarray}
where $P_{{\rm ec},m}(q )$ is the form factor of the micelle formed by
the aggregation of $m$ {{\rm P80}} molecules of which $m_0$ are
involved in the formation of two end-caps and the remaining $m-m_0$
are forming a cylindrical region between them and $p(m)$ represents
the probability density of having micelles with $m$ molecules,
\begin{eqnarray}
p(m)=\frac{C_m}{ \int_{m_0}^\infty C_m \,  dm }.
\label{penne}
\end{eqnarray}
An expression similar to Eq.~3 can be derived for the average
amplitude of polydisperse micelles,
\begin{eqnarray}
  P_{{\rm ec}}^{(1)}(q )&=&\int_{m_0}^\infty p(m) P_{{\rm ec},m}^{(1)}(q ) dm
\label{PP1}
\end{eqnarray}
We have adopted one of the most suitable SAS models for describing
this kind of micellar shape~\cite{amani2011}, which is the one
developed by Kaya~\cite{kaya2004}, here extended to the presence of an
inner hydrophobic domain ({2}-domain) and an outer hydrated polar head
domain ({1}-domain). A representation of this model is shown in
Fig.~2.
\begin{figure}
\centering\includegraphics[width=0.5\textwidth]{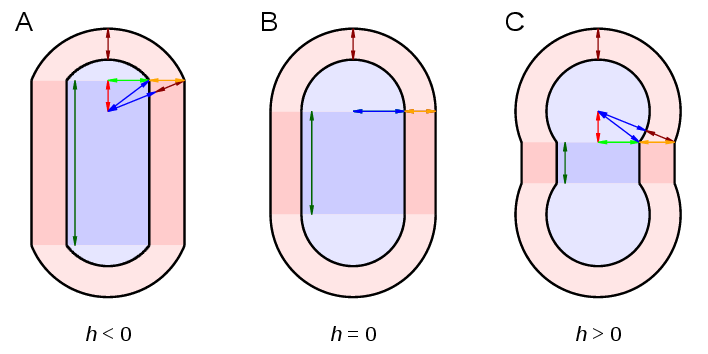}
\caption{Sketches of the globular end-cap cylinder model, with
  negative, null and positive parameter $h$, panels A, B and C,
  respectively. The geometrical parameters of the model represented
  with colored segments, are: $h$ (red, absolute value), $R_{{2},{{\rm
        cyl}}}$ (green), $R_{{2},{{\rm cap}}}$ (blue), $\delta_{{{\rm
        cyl}}}$ (orange), $\delta_{{{\rm cap}}}$ (dark-red) and $L$
  (dark green). Areas with less intense and more intense shadings
  represent the {ED} of end-cap and cylindrical regions, respectively,
  red and blue shading being the corresponding hydrated polar head and
  hydrophobic domains, respectively.}
\label{kayamodel}
\end{figure}
The geometrical parameters of the model are the radius $R_{{2},{{\rm
      cyl}}}$ (green segment, Fig.~2) of the inner
cylindrical domain (intense blue shadow, Fig.~2) and its
length $L$ (dark-green segment, Fig.~2), the thickness
$\delta_{{{\rm cyl}}}$ (orange segment, Fig.~2) of the
outer cylindrical shell (intense red shadow, Fig.~2),
the radius $R_{{2},{{\rm cap}}}$ (blue segment, Fig.~2)
of the inner spherical cap domain (less intense blue shadow,
Fig.~2) and the thickness $\delta_{{{\rm cap}}}$
(dark-red segment, Fig.~2) of the outer shell of the
spherical cap domain. The parameter $h=\pm\sqrt{R_{{2},{{\rm
        cap}}}^2-R_{{2},{{\rm cyl}}}^2}$ (red segment,
Fig.~2) could be negative, null, or positive, as
highlighted in panels A-C of Fig.~2, respectively. To
note, the condition $R_{{2},{{\rm cap}}}>R_{{2},{{\rm cyl}}}$ should
be respected. By observing the right triangles with the colored sides
that appear in panels~A and C of Fig.~2, it is evident
that $\delta_{{{\rm cyl}}}+R_{{2},{{\rm cyl}}}=\sqrt{(R_{{2},{{\rm
        cap}}}+\delta_{{{\rm cap}}})^2-h^2}$, hence only one of two
parameters $\delta_{{{\rm cyl}}}$ and $R_{{2},{{\rm cyl}}}$ can be
considered independent.  The scattering parameters of the model are
the electron densities ({ED}s) of the cylindrical and end-cap regions,
distinguished in hydrated polar domains ($\rho_{1,{\rm cyl}}$ and
$\rho_{1,{\rm cap}}$, more intense and less intense red shadows in
Fig.~2) and in hydrophobic domains ($\rho_{2,{\rm cyl}}$
and $\rho_{2,{\rm cap}}$, more intense and less intense blue shadows
in Fig.~2).  The geometrical parameters of the model can
be related to the aggregation numbers $m$ and $m_0$. Indeed, by
referring to the hydrophobic molecular volume of {{\rm P80}} in the
end-cap domain, $\nu_{{\rm hyd},{{\rm cap}}}$, we have the following
constraint,
\begin{eqnarray}
m_0 \nu_{{\rm hyd},{{\rm cap}}}=\frac{4}{3}\pi R_{{2},{{\rm cap}}}^3\left[1+\frac{3}{2}\frac{h}{R_{{2},{{\rm cap}}}}-\frac{1}{2}\left(\frac{h}{R_{{2},{{\rm cap}}}}\right)^3\right],
\end{eqnarray}
whereas, considering the hydrophobic molecular volume of {{\rm P80}}
in the cylindrical domain, $\nu_{{\rm hyd},{{\rm cyl}}}$, we have
\begin{eqnarray}
  (m-m_0)\nu_{{\rm hyd},{{\rm cyl}}}&=&\pi R_{{2},{{\rm cyl}}}^2 L
\end{eqnarray}
The {X-ray} scattering amplitude $ A_m({\bf q})$, which is defined as
the Fourier transform of the excess {X-ray} scattering length density,
of the core-shell end-cap cylinder formed by $m$ {{\rm P80}}
molecules, derived according to Kaya's model~\cite{kaya2004}, is fully
reported in Eq.~S11 of the {SM} as a function of the components of the
scattering vector ${\bf q}$ parallel and perpendicular to the
cylindrical axis, $q_\| = q \cos \beta_q $ and $q_\perp = q \sin
\beta_q $, respectively ($\beta_q$ is the angle between ${\bf q}$ and
the cylindrical axis and $q$ is the modulus of ${\bf
  q}$). Corresponding orientational integrals $P_{{\rm ec},m}(q )$ and
$P_{{\rm ec},m}^{(1)}(q )$ are defined in Eqs.~S16 and S17 of the
{SM}.  By entering the results of the ladder model,
$C_m=e^{-(\Delta-m_0\delta)/(RT)} e^{-m E} $, where the positive
dimensionless parameter $E=\delta/({{\rm R}}T) -\log C_{\rm P80}-\log
\alpha_1$ has been introduced, we have been able to simplify
Eqs.~3 and 5 according to
\begin{eqnarray}
  P_{{\rm ec}}(q )&=& \int_0^{\pi/2} d \beta_q  \sin \beta_q   \frac{N_2}{D_2},
  \label{PPave}\\
  P_{{\rm ec}}^{(1)}(q )&=& \int_0^{\pi/2} d \beta_q  \sin \beta_q   \frac{N_1}{D_1},
  \label{PP1ave}
\end{eqnarray}
where the working factors $N_2$, $D_2$, $N_1$ and $D_1$ are fully
reported in Eqs.~S18-S21 of the {SM}.  The volumetric properties of
the {{\rm P80}} molecules have been used to calculate all the electron
densities as well as the area per molecule in both end-cap and
cylinder regions. All details are shown in
Sect.~S11.2 of the {SM}.  Besides, the
average number density of the micelles, defined by $n_{\rm ec}=N_A
\int_{m_0}^\infty C_m \, dm$ ($N_A$ is Avogadro's number), is $n_{\rm
  ec}=(N_A/E)e^{-(\Delta-m_0\delta)/({{\rm R}}T)-m_0 E}$.

The effective structure factor $S_M(q)$, a term that reflects the
correlation among micelles, particularly relevant at the high
concentration (particle volume fraction $\eta$ greater than $\approx
0.01$), and that depends on the coupling function $\beta(q)=[P_{{\rm
      ec}}^{(1)}(q )]^2/P_{{\rm ec}}(q )$~\cite{spinozzi2020}, is
modeled with the same approach, based on the HSDY potential, that some
of us have successfully applied to different nanosized
systems~\cite{ortore2009,spinozzi2020,piccinini2022}. In the case of
the nonionic {{\rm P80}} surfactant, the micelle charge is set to
zero. Hence, the only relevant parameters are the effective average
micelle diameter $\sigma_{\rm ec}$ (so that the volume fraction that
appears in the PY expression of $S_0(q)$ (see Eq.~S1 of Piccinini et
al.~\cite{piccinini2022}) is $\eta=n_{\rm ec}\pi\sigma_{\rm ec}^3/6$),
the depth $J$ of the attractive potential's well, and its the decay
range $d$.  Moreover, since our experimental data show a $q^{-4}$
behavior at low $q$ (Sect.~3.3) probably due to the
presence of very large micellar aggregates, the final equation used to
fit all the experimental SAXS {differential macroscopic cross section}
recorded for the samples containing only {{\rm P80}} is
 \begin{eqnarray}
   {\frac{d\Sigma}{d\Omega}}_{{\rm ec}}(q )&=& n_{\rm ec} r_{\rm e}^2 P_{{\rm ec}}(q ) S_M(q)+k_{\rm por}q^{-4}
   \label{eqec}
   \end{eqnarray}
where $k_{\rm por}$ represents the Porod's constant. The factor
$r_{\rm e}=0.28\cdot 10^{-12}$~cm is the scattering length of the
electron.

\paragraph{{SAXS of stacked polydisperse platelets in the form of barrel}}
\label{saxs2}

  \begin{figure}
  \begin{center}
    \includegraphics[width=0.8\textwidth]{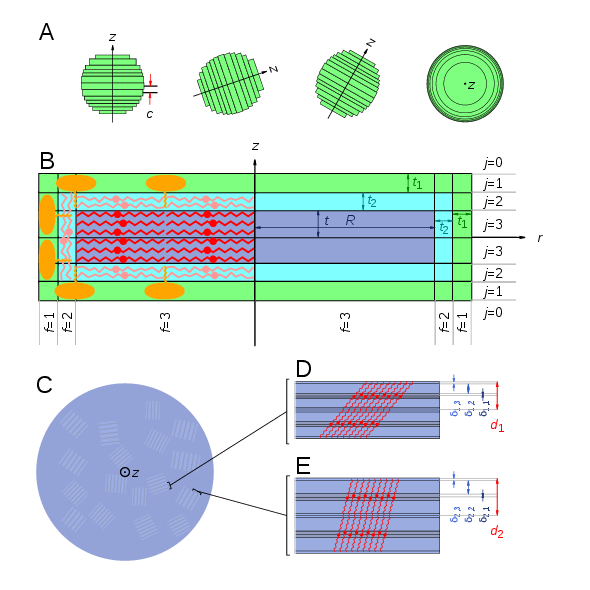}
  \end{center}
  \caption{The platelets model for LNPs. A) Randomly
      oriented, polydispersed, barrel-like LNPs. Each comprises a
      stack of $N_c$ platelets, orthogonal to their main axis $z$,
      separated by a distance $c$ (red marks), with $N_c$ and $c$
      following Gaussian distributions and a cross-section radius
      distributed according to an elliptical profile (Fig.~S2). B)
      Section of a single platelet along its central axis $z$. The
      inner cylinder (blue), made of crystalline CP molecules (red),
      has half-height $t$ and radius $R$ (vertical and horizontal
      dark-blue double arrows, respectively). The $t/R$ ratio is not
      in scale and has been chosen for the sake of visualization.
      Amorphous-CP molecules (pink) form a cylindrical shell (cyan)
      with thickness $t_2$ (dark-cyan double arrows) within another
      cylindrical shell (green) with thickness $t_1$ (dark-green
      double arrows) made of P80 molecules (orange) and
        bound water (not shown). Labels $f = 1$, 2, and 3 mark the
      regions with different ED profiles along $z$. The $f=3$ region
      has six layers ($j = 1$, 2, 3, and those specular to the central
      plane orthogonal to $z$, with thicknesses $t_1$, $t_2$, and $t$,
      respectively) with distinctive EDs. The $f=2$ region has layers
      $j=2$ and 3 with the same ED, while for the $f=1$ region all the
      layers have the same ED. C) Cross section of the
        inner cylinder (outgoing $z$ axis) in which small crystalline
        domains, randomly oriented and containing several parallel CP
        bilayers, are shown.  The crystalline domains are
        distinguished into two groups, with shorter and longer
        lamellar distances. A zoom of two adjacent bilayers for each
        of the two groups of crystalline domains is shown in panels D
        and E, respectively, where the red double arrows represent the
        corresponding lamellar distances $d_1$ and $d_2$.  A
        hypothetical disposal of the CP molecules, oriented
        differently in the two types of bilayers, is shown. The three
        specular layers, shown by decreasing intensities of blue,
        represent the {ED} of the carboxyl group, the middle and
        terminal chains, respectively. Their corresponding thicknesses
        $\delta_{k,1}$, $\delta_{k,2}$ and $\delta_{k,3}$
          ($k=1,2$) are indicated by the double arrows
        shown on the right with the same color as the {ED}
        domains.
    }
\label{platmod}
\end{figure}

SAXS data of {{\rm CP}} solid LNPs stabilized by {{\rm P80}}
(Sect.~3.3) show two sets of
low-order diffraction peaks, similar to previous
results~\cite{gordillogaleano2018,Barbosa2018,blasi2007,blasi2013,deSouza2012,blasi2011,blasi2013b},
that grow over a typical bilayer band, widely seen in SAXS experiments
of flat bilayers~\cite{FeiginSvergun1987}, suggesting the presence of
platelets.  The first peak's positions of the two families are at
$\approx 0.160$~{\AA}$^{-1}$ and $\approx 0.143$~{\AA}$^{-1}$,
corresponding to repeat distances of $\approx 39.3$~{\AA} and $\approx
43.9$~{\AA}, respectively.  The low $q$ behavior of these SAXS
curves show a power trend $q^{-p}$ with
exponent $2<p<3$, far from the characteristic $p=2$ value of freely
rotating platelets in solution~\cite{kratky1982}, indicating that a
certain degree of parallel platelet-platelet stacking interaction
would occur.  Based on these preliminary observations, together with
the AFM and DLS results shown in Sect.~3 and by taking into
account the detailed model developed by Schmiele et
al.~\cite{Schmiele2013,Schmiele2015} for platelet systems, we have
worked out a novel model aimed at analyzing SAXS curves in the whole
$q$-range.  As a matter of fact, it should be noted that no model
among those reported in the literature has proved capable of fitting
the SAXS curves in the entire range of $q$.  This novel model
(Fig.~3A) is based on the
following assumptions.
\begin{enumerate}[label={(\roman*)}, align=left,wide=0pt,itemsep=0.5 pt]
\item
  \label{start}

The platelets are composed of three cylindrical structures
  that are embedded inside one another (Fig.~3B). The
  innermost cylinder (blue,
  Fig.~3C) is made up of lamellar
  layers consisting of CP molecules (red) with internal structures
  (Fig.~3D and E). This cylinder is surrounded by a
  cylindrical shell (cyan) consisting of widely-spaced P80 aliphatic
  chains (orange) and amorphous CP molecules (pink). The outer
  cylinder (green) is made up of P80 polar heads and bound water
  molecules that bridge the hydrophilic moieties, as illustrated in
  atomistic simulations of phospholipid
  membranes~\cite{Calero2019}.

\item
  \label{poly1}
The innermost cylinder (shown in Fig.~3 in blue
color) has both the radius $R$ and the height $2{t}$
polydisperse. The mean value of the radius is indicated with $R_0 =
<\!R\!>$, its dispersion index is $ \xi_R =(<\!R^2\!>-R_0^2)^{1/2}/R_0$,
while $t_0=<\!{t}\!>$ indicates the mean value of half the
height of the inner cylinder, with dispersion index $ \xi_{t}
=(<\!{t}^2\!>-t_0^2)^{1/2}/t_0$.
\item
  \label{polyR}
  According to AFM results, we assume that LNPs are barrel-shaped
  particles, defined by a maximum and a minimum radius of the circular
  cross-section, $R_{M}$ and $R_m=\nu R_{M}$, respectively, with
  $0<\nu<1$, $\nu$ being the ``bulging'' parameter of the
  barrel. Also, we assume a smooth variation of the barrel's circular
  cross-section radius according to an elliptical profile, as depicted
  in Fig.~S2 of the {SM}, where we have also plotted the theoretical
  distribution function of the barrel circular cross-section radius,
\begin{eqnarray}
  p(R,R_M,R_m)=\left\{\begin{array}{ll}\frac{R-R_m}{(R_M-R_m)\sqrt{(R_M-R)(R_M+R-2 R_m)}} & R_m\leq R < R_M\\
    0 &\mbox{otherwise}
    \end{array}
    \right.
  \label{defpr}
\end{eqnarray}
According to this view, the barrel shape is obtained by the stacking
of parallel cylindrical platelets
(Fig.~3A). Since AFM results
indicate a polydispersion of the barrel size, we assume a Gaussian
distribution of the maximum circular cross-section radius $R_M$ of the
barrel, centered at $R_{M,{\rm max}}$ and with standard deviation
$\xi_{R_M} R_{M,{\rm max}}$, As a consequence, the overall
distribution function $p(R)$ of the platelet radius is written as,
\begin{eqnarray}
  p(R)&=&\int_{R_{M,\rm lb}}^{R_{M,\rm ub}} p(R,R_M,\nu R_M) \, p(R_M)\,dR_M \\
 p(R_M)&=& \frac{1}{Z_{R_M} }e^{-(R_M-R_{M,{\rm max}})^2/(2 \xi_{R_M}^2 R_{M,{\rm max}}^2)}
  \label{ebarrel}
\end{eqnarray}
The lower and the upper bounds of the integral are $R_{M,\rm lb}={\rm
  max}\{R_{M,{\rm max}}(1-p_G\xi_{R_M}),R_{M,{\rm min}}\}$ and
$R_{M,\rm ub}=R_{M,{\rm max}}(1+p_G \xi_{R_M} )$, respectively, where
$p_G\approx 3$ represents the number of standard deviations of the
Gaussian taken into consideration, whereas $R_{M,{\rm min}}$
represents the minimum value of $R_{M}$, a parameter necessary in
order to avoid nonphysical negative values of $R_{M}$. The
normalization factor, $ Z_{R_M} $, can be analytically calculated as
reported in Eq.~S22 of the {SM}.  Examples of $ p(R)$ calculated with
Eq.~13 are reported in Fig.~S3 of the {SM}. To note, the
average platelet radius and its dispersion are calculated according to

  \begin{eqnarray}
  R_0 &=&    \int_{\nu R_{M,\rm lb}}^{R_{M,\rm ub}} \, R \, p(R) dR
  \label{er0}
\\
  \xi_R^2  &=&  \frac{1}{R_0^2}\int_{\nu R_{M,\rm lb}}^{R_{M,\rm ub}} (R-R_0)^2 \,p(R)\, dR
  \label{er2}
\end{eqnarray}
with $\int_{\nu R_{M,\rm lb}}^{R_{M,\rm ub}} \,p(R) \,dR=1$. Analytical
expressions of $R_0$ and $\xi_R$ are given in Eqs.~S23-S24 of the {SM}.

\item
  \label{polyt}
A Gaussian function also describes the distribution function of $t$
with the maximum at the position $t_{\rm max}$ and the standard
deviation defined as $\xi_{t_{\rm max}} t_{\rm max}$. Since $t$ is a
positively defined quantity, the average thickness, $t_0$ and the
dispersion, $\xi_{t}$, are calculated by integrating the Gaussian
function only in a positive range of $t$, as described in detail in
the Sect.~S5 of the {SM}.  To note, $t$ is
represented by a dark-blue arrow in
Fig.~3B.
\item
\label{large}
Platelets are highly anisometric cylinders, with $R \gg t $, hence,
according to scattering theory~\cite{kratky1982}, the SAXS signal only
depends on the excess {ED} along the axis $z$ of the platelet (drawn
in Fig.~3B). As a consequence, there are three distinct {ED} profiles
along $z$, indicated with the label $f=1,2,3$, each of them formed by
$3$ specular layers in respect to the middle plane orthogonal to the
$z$ axis (Fig.~3B). Such layers are indexed by $j=1,2,3$, and the
corresponding thicknesses are $t_1$, $t_2$ and $t$.  In
  positions labeled with $j=0$ in Fig.~3B there are stacked platelets
  with their $t_1$, $t_2$ and $t$ layers, with a stacking interlayer
  distance ${\Delta t}\simeq 1 $\AA, as discussed in section 3.3.2
  (Fig. 9M). We associate ${\Delta t}/2$ to each stacked platelet as a
  correction to the thickness $t_1$ of the hydrated-P80
  layer. For the {ED} profile indexed with $f=3$, the
$3$ layers have distinct values of {ED}, as shown in Fig.~3 with
{blue} ($j=3$), cyan ($j=2$) and green ($j=1$) colors. Differently,
for the {ED} profile with index $f=2$, two layers have the same {ED},
shown in cyan ($j=2,3$) in Fig.~3B, whereas for $f=1$ all layers have
the same {ED}.  We also assume smooth transitions of {ED}s from two
subsequent layers and from the last layer to bulk water by adopting
the error function to describe the smooth effect (see Fig.~S4 of the
{SM} and Spinozzi et al.~\cite{Spinozzi2010} for details). The smooth
parameter from $j$-layer to $(j-1)$-layer is the standard deviation
$\sigma_{{\rm pl},j}$ of the error function.
\item
  \label{para1}
 The stacking among roughly parallel platelets
 (Fig.~3A) is described by the
 para-crystal theory applied along the $z$ direction, with a repeat
 distance $c=2(t+t_2+t_1+{\Delta t}/2)$
 (Fig.~3A, red arrows) and
 distortion parameter $g_c=\sigma_c/c$, $\sigma_c$ being the standard
 deviation of $c$. The number of stacking platelets, $N_c$, is
 polydisperse, according to a Gaussian distribution function
 $p_{N_c}(N_c)$, with the maximum at the position $N_{c,\rm max}$ and
 the standard deviation indicated with $\sigma_{N_c}$. Since $N_c$
 cannot be negative, the average para-crystal structure factors, as
 well as the average number of platelets, $N_{c,0}$, are calculated by
 integrating the Gaussian distribution function only in a positive
 range of $N_c$, as detailed in the Sect.~S6 of the
   {SM}.
\item
  \label{para2}
The {{\rm CP}} molecules in the innermost cylinder (Fig.~3C) are
organized into three groups, two of which correspond to two nano-sized
lamellar domains (Fig.~3D and E) and a the third group forming an
amorphous domain. The molar fraction of {{\rm CP}} in the three groups
are named $y_k $, with the obvious condition $\sum_{k =1}^3y_k
=1$. The lamellar orders of the domains ({{\rm ld}}) are described by
the para-crystal scheme of Fr{\"u}hwirth et al.~\cite{Fruehwirth2004},
defined by the repetition distance $d_k $ (Fig.~3D and E, red arrows),
the distortion $g_{{\rm ld},k }$ and the average repeat number
$N_{{\rm ld},k }$, with $k =1,2$. In turn, the repetition distance is
$d_k =2(\delta_{k ,1}+\delta_{k ,2}+\delta_{k ,3})$, where $\delta_{k
  ,i} $ is the thickness of the $i $-layer of {ED} corresponding to
the carboxyl group, the middle, and the terminal chains of the {{\rm
    CP}} molecules, with $i =1,2,3$, respectively (see arrows with
decreasing intensity of blue in Fig.~3D and E).  Smooth transitions
from $i$-layer to $(i-1)$-layer are modeled based on the error
function with standard deviation $\sigma_{k,i}$. To note, the
$0$-layer has the {ED} corresponding to the average of the {ED}s of
the three layers, as shown in Fig.~S5 and
Eq.~S61 of the {SM} and Ref. Spinozzi et al.~\cite{Spinozzi2010}.

\item
  \label{final}
The {{\rm P80}} molecules are divided into two groups. Those in the
first group (with molar fraction $y_{\rm P80}$) are distributed on the
platelets' surface, with their large polar head in the layer $j=1$ and
their hydrophobic chain in the intermediate layer ($j=2$), among the
{{\rm CP}} molecules considered in amorphous configuration (Fig.~3B).
The polar heads of {\rm P80} molecules are hydrated by
  bound water in the $j=1$ layer. As discussed in
  Sect.~3.3.2, the bound water is responsible
  for the collapse of the platelets.  The second group of {{\rm P80}}
  molecules, with a molar fraction $1-y_{\rm P80}$, consists of all the
  molecules forming end-cap cylinder micelles, according to the model
  described in the Sect.~2.2.4, paragraph ``{SAXS of growing and
    interacting end-capped cylindrical micelles}''. However, as
  discussed in Sect.~3.3, we found that $y_{\rm P80}=1$. Hence, there
  are no end-cap cylinder micelles in the LNPs, although the general
  theory includes them.
\item
  \label{final2}
  Both the height of the barrel, $H=c\,N_c$, and the circular
  cross-section radius of the barrel, $R$, are considered larger than
  $\approx 1/q_{\rm min}$, $q_{\rm min}$ being the minimum modulus of
  the scattering vector detectable by SAXS experiments. Hence, the
  contribution of the whole barrel to the SAXS signal depends on the
  average surface of the barrel and is due to the excess {ED} of all
  the molecules within the barrel ({{\rm CP}}, {{\rm P80}} and bound water
  between the platelets) with respect to the bulk water, a case similar
  to that described by Porod's law.  To note, in the case of a
  barrel-like LNP that interacts with other molecules, such as
  proteins, the SAXS contribution of the barrel surface will be
  approximated by the form factor of $N_s$ layers of different {ED}s
  in planar geometry, with possible smooth transitions, according to
  the classical scattering theory~\cite{kratky1982}.  The distribution
  function of $H$ corresponds to $p(H)=(1/c)p_{N_c}(H/c)$. We have
  also developed a simple Monte Carlo method to derive the
  distribution function of the center-to-border distance of the
  barrel, $p(R_{\rm c})$, by combining the distributions functions
  $p(R_M)$ and $p(H)$. Details are given in the
  Sect.~S7 of the
    {SM}.
\end{enumerate}

We will now derive the SAXS {differential macroscopic cross section}
of platelets according to all these assumptions, from {{(i) to
{{(ix).  According to scattering theory~\cite{kratky1982}, the
SAXS {differential macroscopic cross section} of flat ({{\rm fl}}),
thin and not interacting ({{\rm ni}}) platelet with surface ${S_{\rm
    fl}}$ and number density $n_{\rm pl}$ is
\begin{eqnarray}
  {\frac{d\Sigma}{d\Omega}}_{{\rm fl},{\rm ni}}(q)= n_{\rm pl} r_{\rm e}^2 \frac{2\pi}{q^2}{S_{\rm fl}} |A_{\rm fl}(q)|^2
\end{eqnarray}
where $A_{\rm fl}(q)=\int \delta \rho(z)e^{iqz}dz$ is the Fourier
transform the excess {ED} profile $\delta \rho(z)$ along the direction
$z$ perpendicular to the platelet.  The number density of platelets,
$n_{\rm pl}$, with inner radius $R$, half-inner length $t$ and shell
thicknesses $t_1$ and $t_2$ can be calculated considering the {{\rm
    CP}} w/v concentration, $c_{{\rm CP}}$, by $n_{\rm pl}=N_A c_{{\rm
    CP}}/(M_{\rm CP} N_{{\rm CP},{\rm pl}})$, where $M_{\rm CP}$ is
the {{\rm CP}} molecular weight and $N_{{\rm CP},{\rm pl}}$ is the
number of {{\rm CP}} molecules in the platelet, which can be derived
on the basis of the mass balance, as shown in Eq.~S105 of the {SM}.
By referring to assumption~{{(v), since for platelets we have
three {ED} profiles along $z$ ($f=1,2,3$), the {differential
  macroscopic cross section} of the platelets is
\begin{eqnarray}
  {\frac{d\Sigma}{d\Omega}}_{{\rm pl},{\rm ni}}(q)&=&n_{\rm pl} r_{\rm e}^2 \frac{2\pi}{q^2}\left[
        \pi ((R+t_2+t_1)^2 -(R+t_2)^2) A_{{\rm fl},1}^2(q)
    \right .\nonumber \\&&\left. +    \pi ((R+t_2)^2 -R^2) A_{{\rm fl},2}^2(q)
   + \pi R^2 A_{{\rm fl},3}^2(q)
    \right] \label{em}
\end{eqnarray}
The real functions $A_{{\rm fl},f}(q)$ for specular layers with smooth
transitions based on error functions are reported in Eq.~S44 of the
{SM}.  By substituting the expression of $N_{{\rm CP},{\rm pl}}$ shown
in Eq.~S105 of the {SM} and considering both the polydispersion model
described in assumption~{{(ii) and the stacking correlation
described in assumption~{{(vi), the {differential macroscopic
  cross section} of interacting ({{\rm in}}) polydisperse platelets,
averaged over $R$ and $t$ ({{\rm av}}), is
\begin{eqnarray}
  {\frac{d\Sigma}{d\Omega}}_{{\rm pl},{\rm in},{\rm av}}(q)&=& r_{\rm e}^2
\phi_{\rm CP}\left( 1 +     \frac{\nu_{{\rm P80},{\rm hyd}}y_{\rm P80}}{ {\bar \nu}_{\rm CP}r_{{\rm CP},{\rm P80}}     k_{r_{{\rm CP},{\rm P80}} } } \right)
\nonumber \\&&\times \frac{\pi}{q^2}\left(
<\! t_1(t_1+2(R+t_2))( R + t_2)  ^ {-2}\!>_R
<( t + t_2 )^{-1}A_{{\rm fl},1}^2(q)>_t
\right.\nonumber \\&& \left.
+<\!   t_2(t_2+2R)( R + t_2)  ^ {-2}\!>_R
     <( t + t_2 )^{-1}A_{{\rm fl},2}^2(q)>_t
\right.\nonumber \\&& \left.
+<\! R^2( R + t_2)  ^ {-2}\!>_R
     < ( t + t_2 )^{-1}A_{{\rm fl},3}^2(q)>_t
 \right)S_{{\rm pl}}(q)
 \label{placorr}
  \end{eqnarray}
where $\phi_{\rm CP}=N_A c_{{\rm CP}} {\bar \nu}_{\rm CP}/M_{\rm CP} $
(see Eq.~S109 of the {SM}) is the overall {{\rm CP}} volume fraction,
$r_{{\rm CP},{\rm P80}} $ is the nominal molar ratio between {{\rm
    CP}} and {{\rm P80}} molecules, with an eventual correction factor
and $ k_{r_{{\rm CP},{\rm P80}} }$. The terms $\nu_{{\rm P80},{\rm
    hyd}}$ and ${\bar \nu}_{\rm CP}$ are the volumes of the
hydrophobic tail of {{\rm P80}} and the mean volume of {{\rm CP}},
as detailed in Eq.~S66 and S108 of the {SM},
  respectively. The radial averages ($<\dots>_R$) are
calculated on the basis of the function $p(R)$, as shown in Eq.~S52 of
the {SM}.  The $t$-averages $<\!(t+t_2)^{-1 }A_{{\rm fl},f}^2(q)\!>_t$
are determined as fully described in Eq.~S47 of the {SM}.  The factor
$S_{{\rm pl}}(q)$ in Eq.~18 represents the
platelet-platelet structure factor, which is calculated according to
para-crystal order along the $z$ direction. The expressions can be
found in literature~\cite{Fruehwirth2004, Spinozzi2010}.

Regarding the SAXS {differential macroscopic cross section}s of the
two groups of randomly oriented nano-sized lamellar domains ({{\rm
    ld}}) of {{\rm CP}} within the inner cylinder, foreseen by
assumption~{{(vii), according to scattering theory it can be shown
that they are two terms that add up to the one due to the platelets
since the average cross-terms between the cylindrical layer of the
platelets, and the nano-domains drop to zero.  Considering the stacks
of flat bilayers, as shown in Fig.~3D and
  E, their {differential macroscopic cross section} is
\begin{eqnarray}
  {\frac{d\Sigma}{d\Omega}}_{{\rm ld},{\rm in}}(q)= r_{\rm e}^2
       \phi_{{\rm CP},3} \frac{2\pi}{q^2}
       \sum_{k=1}^2 \frac{y_k}{d_k}
       A_{{\rm ld},k}^2(q)S_{{\rm ld},k}(q)
  \label{eqdom}
\end{eqnarray}
where $\phi_{{\rm CP},3}$ is the volume fraction of {{\rm CP}} in the
inner region of the platelets (see Eq.~S110 of the
  {SM}) and $A_{{\rm ld},k}(q)$ is the Fourier transform
of the excess {ED} profile of the 3-specular layers of the
$k$-nano-domain calculated with respect to the average {ED} of the
{{\rm CP}} molecules (represented in {blue}
in Fig.~3). Its expression is given in Eq.~S62 of the
{SM}. The stacking between {{\rm CP}} $k$-domains is described by the
para-crystal structure factor $S_{{\rm ld},k}(q)$.

The SAXS contribution due to the overall barrel-like surface, based on
the assumption {{(ix), is
\begin{eqnarray}
  {\frac{d\Sigma}{d\Omega}}_{\rm brl}(q)= n_{\rm brl} r_{\rm e}^2 \frac{2\pi}{q^2}<\!S_{\rm brl}\!> |A_{\rm brl}(q)|^2
  \label{saxsbs}
\end{eqnarray}
where $n_{\rm brl}$ is the average number density of barrels (see
Eq.~S123 of the {SM}), $<\!S_{\rm brl}\!>$ is the average barrel
surface (calculated according to Eq.~S132 of the {SM}) and $A_{\rm
  brl}(q)$ is the Fourier transform of the excess {ED} profile along
the direction perpendicular to the barrel surface, fully described in
Eq.~S138 of the {SM}. It can be easily shown that the scattering
cross-term between the barrel and the platelets has a mean value that
tends to be zero.

The final equation used to fit the SAXS data of LNP samples, which
includes all the assumptions {{(i)-{{(ix) is the sum of
Eqs.~18, 19, 20 and 10,
\begin{eqnarray}
  {\frac{d\Sigma}{d\Omega}}_{\rm LNP}(q)= {\frac{d\Sigma}{d\Omega}}_{{\rm pl},{\rm in},{\rm av}}(q)+ {\frac{d\Sigma}{d\Omega}}_{{\rm ld},{\rm in}}(q)+  {\frac{d\Sigma}{d\Omega}}_{\rm brl}(q)+ {\frac{d\Sigma}{d\Omega}}_{{\rm ec}}(q)
  \label{fitcp}
\end{eqnarray}
where in the term ${\frac{d\Sigma}{d\Omega}}_{{\rm ec}}(q)$
(Eq.~10), which accounts for the SAXS contribution of {{\rm
    P80}} molecules that are not involved in the platelets, the number
density of end-cap cylindrical micelles, $n_{\rm ec}$, is calculated
as widely described in Sec.~2.2.4, paragraph ``{SAXS
    of growing and interacting end-capped cylindrical
    micelles}'', and considering the available molar
concentration of {{\rm P80}} as large as $C_{\rm P80}=c_{\rm
  P80}(1-y_{\rm P80})\,d_{\rm wat}\,/M_{{\rm P80}}$.

\paragraph{{Global-fit}}
\label{gb}
Considering the interplay between the SAXS models introduced in
Sect.~2.2.4, paragraphs ``{SAXS of growing and
    interacting end-capped cylindrical micelles}'' and ``{SAXS of
    stacked polydisperse platelets in the form of
    barrel}'', all SAXS curves of samples containing
only {{\rm P80}} and samples of LNP (containing both {{\rm P80}} and
{{\rm CP}}) can be analyzed by a unique optimization procedure,
referred to as global-fit~\cite{spinozzi2014}.  All model parameters
are divided into two classes: the first-class includes the common
parameters, such as the volumes of chemical groups, which are
optimized to a single value for all curves; the second-class includes
single-curve parameters, which can assume an independent value for
each curve.  The merit function to be minimized is
\begin{eqnarray}
  {\cal H}&=&\chi^2+\alpha \, L.
  \label{merit}
\end{eqnarray}
In this equation, the term $\chi^2$ is the standard reduced chi-square
of all the $N_v$ experimental SAXS curves,
\begin{eqnarray}
  \chi^2&=&\frac{1}{N_v}\sum_{v=1}^{N_v} \frac{1}{N_{q,v}}
\sum_{j=1}^{N_{q,v}}
\left(\frac{\frac{d\Sigma}{d\Omega}_{v,{\rm ex}}(q_j)-
  \frac{d\Sigma}{d\Omega}_{v,{\rm th}}(q_j)}{\sigma_v(q_j)}\right)^2,
\end{eqnarray}
where $N_{q,v}$ is the number of $q$-points of the $v$-curve,
$\frac{d\Sigma}{d\Omega}_{v,{\rm ex}}(q_j)$, $\sigma_v(q_j)$ is the
experimental standard deviation and $\frac{d\Sigma}{d\Omega}_{v,{\rm
    th}}(q_j)$ is the fitting curve calculated based on either
Eq.~10 or Eq~21, depending on the kind of sample
($s={\rm P80}$ or $s={\rm LNP}$).  The second term, $L$, is the
regularization factor aimed to reduce unlikely oscillations of
single-curve parameters related to samples with the closest
chemical-physical conditions (composition, concentration, and
temperature). The term $L$ is indeed defined by
\begin{eqnarray}
  L&=&
  \sum_{s={\rm P80},{\rm LNP}}
  \sum_{p=1}^{N_{p,s}} \sum_{v=1}^{N_v}\left(1-\frac{X_{p,v}}{X_{p,v^\prime}}\right)^2,
  \label{eqreg}
\end{eqnarray}
where the index $s$ in the first sum distinguishes the kind of sample,
the index $ p$ is the label of the $ p^{\rm th}$ of the $N_{p,s}$
single-curve fitting parameters, $X_{p,v}$ is the value of the
parameter used to fit the $v$-curve and $X_{p,v^\prime}$ is the value
of the parameter used to fit the $v^\prime$ curve, which is one of the
sample with the closest chemical-physical conditions to the sample of
the $v$-curve. The closest curve is the one that minimizes the term
$(1-C_v/C_{v^{\prime}})^2+(1-T_v/T_{v^{\prime}})^2$, where $C_v$ is
the concentration of either {{\rm P80}} or {{\rm CP}}.  The
minimization of the merit function and the evaluation of the
uncertainties of fitting parameters are achieved according to a
combination of Simulated Annealing and Simplex methods, as detailed
described by Moretti et al.~\cite{Moretti2020}. The constant $\alpha$
in Eq.~22 is fixed to ensure that, at the end of the
minimization, the factor $\alpha\,L$ does not overcome $\approx 10\%$
of the merit function ${\cal H}$.  The present model has been
integrated into the GENFIT software~\cite{spinozzi2014}.

\section{Results and discussion}
\label{rad}

\subsection{{DLS}}
\label{sdls}

In Fig.~4A, we report the second-order intensity
autocorrelation functions of a solid LNP dispersion measured on
different days after it was prepared. These functions exhibit a single
exponential decay, suggesting that the sample primarily comprises
particles of the same size. Accordingly, we analyzed the data with a
single Gaussian distribution function of the LNP's hydrodynamic
radius, $R_H$, as in Eq.~S9 of the {SM}, and achieved optimal fits
(black solid lines in Fig.~4A, with fitting parameters in
Table~S2 of the {SM}).  We show the resulting hydrodynamic radius
distributions in Fig.~4B, and the histogram of the average
values $<\!R_H\!>$ as a function of time from sample preparation in
Fig.~4C. These results indicate that the LNP
  size is stable, with an average hydrodynamic radius of
  $<\!R_H\!>\approx 950$~{\AA}, with rather limited temporal
  variations (in the order of 2\%), and with a dispersion index
  $\xi_{R_H}\approx 0.3$. Additionally, we found that
the particles were slightly negative with a $\zeta$ potential of $-6.5
\pm 0.6$~mV, which remained relatively constant throughout the
investigation.

\begin{figure}
\begin{center}
  \includegraphics[width=\textwidth]{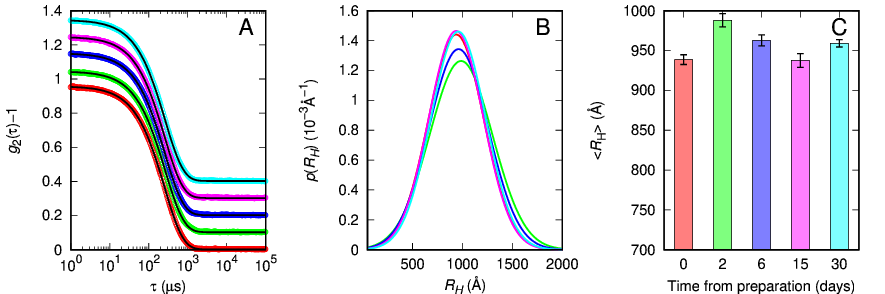}
  \caption{DLS results of LNP. A) Auto-correlation functions of LNP
    recorded at 0 (red), 2 (green), 6 (blue), 15 (magenta) and 30
    (cyan) days after the sample preparation with corresponding best
    fits (solid black lines). A factor of $0.1$ vertically scales data
    for clarity. B) Distribution functions of the hydrodynamic radius
    $R_H$ obtained by the the best fit of DLS auto-correlation
    functions. C) Mean particle hydrodynamic radius of the
    distributions shown in panel~B. The color code used in
      panel~B and C is as in panel~A.}
\label{figdls}
\end{center}
\end{figure}

\subsection{{AFM}}

\label{safm}

In Fig.~5A, we show a representative image of tens
non-contact mode AFM observations obtained from a solid LNP dispersion
diluted to 0.1~{g/L}. The particles show an elongated, barrel-like
shape and are noticeably polydisperse. In Fig.~5B-I, we
show magnifications of the images centered on the single
particles. The jagged morphology in the region close to the
  border of AFM images (e.g., Fig.~5G and I)
  is consistent, at least to some extent,
    with the presence of an internal structure formed by
  parallel sheets. As described in the Sect.~2,
we determined the distribution function of the center-to-border
distance $R_{\rm c}$, along different straight lines passing from the
center, of $\sim 50$ particles directly observed by AFM
(Fig.~5J). The mean value of $R_{\rm c}$ is $\approx
912$~{\AA}, with a dispersion as large as $0.1$, in excellent
agreement with the mean hydrodynamic radius $<\!R_H\!>$ measured by
DLS.

\begin{figure}
\begin{center}
  \includegraphics [width=0.65\textwidth]{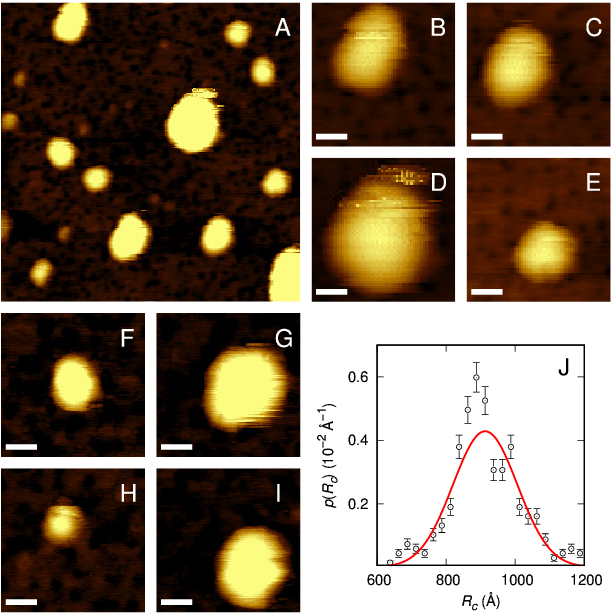}
  \caption{AFM results for LNPs. Panel~A: example image of a sample
    recorded in non-contact mode. Panels~B-I: square
      magnifications of single LNPs. The bottom horizontal bars span
      1000~{\AA}. Panel~J: distribution function of the
    center-to-border distance obtained with ImageJ
    software~\cite{Schneider2012} by selecting 300 distances measured
    in random directions passing through the center of 50 individual
    LNPs. The grid size was $50$~{\AA}, and the error bars were
    assigned according to Poisson statistics. The solid red line
    represents the best fit through a Gaussian, with the center at
    $912\pm 6$~{\AA} and dispersion $0.100\pm 0.006$.}
\label{afmtotal}
\end{center}
\end{figure}

\subsection{SAXS}
\label{ssaxs}

\begin{figure}
\begin{center}
  \includegraphics[width=0.75\textwidth]{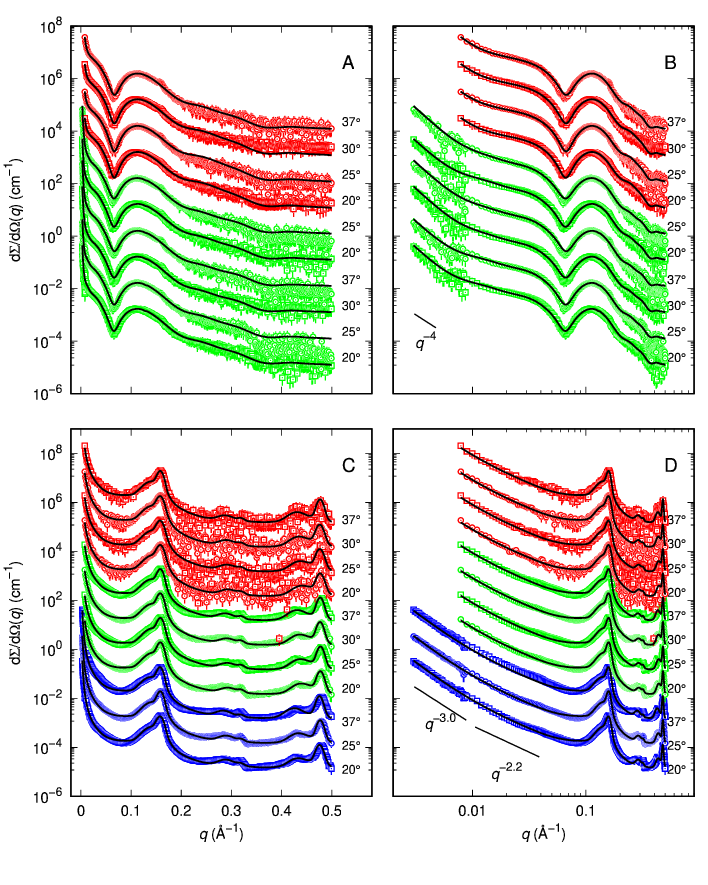}
\caption{Synchrotron SAXS curves recorded at the ID02 beamline at ESRF
  for {{\rm P80}} (panels A-B) and LNP (panels~C-D) samples reported
  in semi-logarithmic plots (panels A and C) and in logarithmic plots
  (panels B and D), respectively. For a better visualization, curves
  have been stacked by multiplying for a factor $10^{m-1}$, where $m$
  is the index of the row from the bottom.  In panels A-B, red and
  green points refer to 13.3 and 1.7~{g/L} {{\rm P80}} concentration,
  respectively. In panels C-D, red, green, and blue points refer to
  80.0, 40.0, and 1.0~{g/L} LNP concentration, respectively. Solid
  black lines are the best fits obtained with the global-fit method.}
  \label{SAXSfit}
\end{center}
\end{figure}

SAXS curves recorded at the ID02 beamline of ESRF for water dispersion
for {{\rm P80}} in different concentrations and temperatures are shown
in Fig.~6A (semi-logarithmic plot) and B (logarithmic
plot), whereas SAXS curves of solid LNPs formed by {{\rm CP}} and
stabilized by {{\rm P80}} are presented in Fig.~6C
(semi-logarithmic plot) and D (logarithmic plot).  We show in Fig.~S7
of the {SM} additional SAXS curves, recorded at Austrian SAXS beamline
of ELETTRA, for {{\rm P80}} and solid LNPs samples of a second
preparation batch but in a more limited number of conditions in terms
of concentration and temperature.  We analyzed simultaneously by a
unique calculation, according to the SAXS models fully described in
the Sect.~2.2.4, all the {{\rm P80}} and the LNP sets of curves
recorded at ESRF and ELETTRA.  We find that the results for each
experimental campaign are very similar. Therefore, we present and
discuss here only those from ESRF, while those from ELETTRA are
included in the {SM}.

First, to reduce the number of free fitting-parameters, we duly
exploited all the information related to the composition of the
molecules {{\rm P80}} and {{\rm CP}}, including the volume of the
different chemical groups and their dependence on the temperature, to
calculate the electron densities of the domains in each of the regions
that constitute the end-capped cylindrical micelles (Sect.~S11.2 of
the {SM}) and the platelets (Sect.~S11.3 of the {SM}). The
  parameter $y_{\rm P80}$, which represents, for the LNP samples, the
  mole fraction of P80 bound to the platelets, was always found equal
  to 1, indicating that, for these samples, there are no P80 molecules
  available to form end-cap micelles. To note, the
global fit of 21 SAXS curves was obtained by optimizing 16 first-class
(common) fitting parameters and 280 second-class (single-curve)
fitting parameters (controlled by the regularization method), with an
average of 14 parameters per curve. It should be noted that, despite
the large number of parameters, the validity range of many of them
have been delimited very carefully around known literature values to
ensure physical meaning. The first-class fitting parameters, shared
among all SAXS curves, and their uncertainties are reported in
Table~2.
  \begin{table}
\caption
    {First-class fitting parameters for the SAXS data recorded by the ID02 beamline at ESRF.
  The parameters and symbols are defined in {SM}.
  The unit of
  length and volume are {\AA} and {\AA}$^3$, respectively.
  Validity ranges of fitting parameters:
$^{\rm a}$ $[-1000 , 1000] $ ;
$^{\rm b}$ $[-50 , 50] $ ;
$^{\rm c}$ $[12.0 , 15.0] $;
$^{\rm d}$ $[11.0 , 14.0] $;
$^{\rm e}$ $[14.0 , 17.0] $;
$^{\rm f}$ $[14.0 , 17.0] $;
$^{\rm g}$ $[19.8 , 23.0] $;
$^{\rm h}$ $[26.2 , 27.5] $;
$^{\rm i}$ $[48.0 , 54.0] $;
$^{\rm j}$ $[29.8 , 30.0] $;
$^{\rm k}$ $[0.95 , 1.00] $;
$^{\rm l}$ $[0.95 , 1.00] $;
$^{\rm m}$ $[7.1 , 7.8] $;
$^{\rm n}$ $[0.97 , 1.15] $;
$^{\rm o}$ $[0.97 , 1.15] $;
$^{\rm p}$ $[0.97 , 1.15] $}
\label{fittable1}
\begin{center}

  \begin{tabular}{lllr@{}c@{}lr@{}c@{}l}
    \hline
    $\Delta$
    & (kJ/mol)
  &$^{\rm a}$& $ -352 $ & $\pm$ & $ 4 $  \\
  $\delta$
    & (kJ/mol)
  &$^{\rm b}$& $ -24.8 $ & $\pm$ & $ 0.2 $  \\
  $\nu^\circ_{\rm >C=}$
    & ({\AA}$^3$)
  &$^{\rm c}$& $ 13.0 $ & $\pm$ & $ 0.1 $  \\
  $\nu^\circ_{\rm =O}$
    & ({\AA}$^3$)
  &$^{\rm d}$& $ 12.0 $ & $\pm$ & $ 0.1 $  \\
  $\nu^\circ_{\rm -O-}$
    & ({\AA}$^3$)
  &$^{\rm e}$& $ 16.0 $ & $\pm$ & $ 0.2 $  \\
  $\nu^\circ_{\rm OH}$
    & ({\AA}$^3$)
  &$^{\rm f}$& $ 16.0 $ & $\pm$ & $ 0.2 $  \\
  $\nu^\circ_{\rm CH}$
    & ({\AA}$^3$)
  &$^{\rm g}$& $ 20.9 $ & $\pm$ & $ 0.2 $  \\
  $\nu^\circ_{\rm CH_2}$
    & ({\AA}$^3$)
  &$^{\rm h}$& $ 26.5 $ & $\pm$ & $ 0.3 $  \\
  $\nu^\circ_{\rm CH_3}$
    & ({\AA}$^3$)
  &$^{\rm i}$& $ 50.0 $ & $\pm$ & $ 0.5 $  \\
  $\nu^\circ_{\rm H_2O}$
    & ({\AA}$^3$)
  &$^{\rm j}$& $ 30.0 $ & $\pm$ & $ 0.3 $  \\
  $\beta_{\rm CH_2}$
    &
  &$^{\rm k}$& $ 0.97 $ & $\pm$ & $ 0.01 $  \\
  $\beta_{\rm CH_3}$
    &
  &$^{\rm l}$& $ 1.00 $ & $\pm$ & $ 0.01 $  \\
    $\alpha_{\rm lip}$
&    ($10^{-4}$~K$^{-1}$)
  &$^{\rm m}$& $ 7.20 $ & $\pm$ & $ 0.07 $  \\
    ${\hat d}_{{\rm wat},{\rm cyl}}$
    &
  &$^{\rm n}$& $ 0.99 $ & $\pm$ & $ 0.01 $  \\
  ${\hat d}_{{\rm wat},{\rm cap}}$
    &
  &$^{\rm o}$& $ 1.01 $ & $\pm$ & $ 0.01 $  \\
  ${\hat d}_{{\rm wat},{\rm pl}}$
    &
  &$^{\rm p}$& $ 1.00 $ & $\pm$ & $ 0.01 $  \\
  \hline
  \end{tabular}
\end{center}
      \end{table}

We show the second-class fitting parameters, together with derived
parameters, as a function of temperature for {{\rm P80}} samples
(Fig.~7) and solid LNP samples (Fig.~9).  The
merit function ${\cal H}$ at the end of the minimization, resulted
$6.4$, corresponding to a total reduced $\chi^2$ of $6.1$.

\subsubsection{Polysorbate~80}
The quality of the fits throughout the whole range of $q$ can be
appreciated by observing Fig.~6A and B. Considering the
single-curve fitting parameters (Fig.~7), we first
observe that the inner radius of the two end-caps is $\approx
18$~{\AA} (Fig.~7A), slightly depending on temperature
and concentration. We find that the parameter $h$
(Fig.~7B) is negative, with a value $\approx -10$~{\AA},
and the thickness of the end-cap shell (Fig.~7C) is
around 34~{\AA}. These values, the fitted thermodynamic parameters
$\Delta$ and $\delta$ (Table~2) lead to a very
spheroidal shape of {{\rm P80}} micelles (Fig.~8).

The derived parameter $m_0$ (Fig.~7G), corresponding to
the number of {{\rm P80}} molecules in the end-cap region, is $\approx
14$, very close to the average value $<\!m\!>$ from Eq.~S10 of the
{SM}. Indeed, the probability densities $p(m)$ of finding micelles
with $m$ molecules (Fig.~S11 of the {SM}) is low for
$m>m_0$. Therefore, the cylindrical region of the micelles is almost
negligible, and the micelle in Fig.~8A is the most
representative.

SAXS results for the micelles' size and shape agree well with those
estimated by coarse-grained (MARTINI) Molecular Dynamics (MD)
simulations~\cite{amani2011}.  Nevertheless, the SAXS measure for the
thickness of the {{\rm P80}} hydrophilic shell, made of the very large
three-branched molecule's headgroup, is larger, suggesting a higher
degree of disorder than that estimated by MARTINI.

In particular, the number of bound molecules per polar head in the
end-cap regions is large, $\approx 950$ (Fig.~7H),
corresponding, for the Eq.~S73 of the {SM}, to a hydration level of
$\approx 94\%$ (Fig.~7J), in agreement with SANS
experiments by Nayem et al.~\cite{nayem2020}.  On the contrary, the
number of bound water molecules in the cylindrical region per {{\rm P80}} and the hydration level are much lower, $\approx 110$ and $\approx 66\%$, respectively (Fig.~7I and K).

We observe that the corresponding mass densities, ${\hat d}_{{\rm
    wat},{\rm cap}}$ and ${\hat d}_{{\rm wat},{\rm cyl}}$, of bound water
embedded in the 1-domain of end-cap and cylinder regions are $\approx
1$ (Table~2). Therefore, water near the end-cap areas,
bound to polar heads, has a
density similar to bulk water.

We report the trend of the area per polar head
(Fig.~7L-O), that, as expected, displays differences
among regions and interfaces within the same region.  For example, the
area between 1- and 2-domains in the end-cap region, $\approx
130$~{\AA}$^2$, almost doubles that in the cylinder region, $\approx
62$~{\AA}$^2$, (Fig.~7N and O).

Next, we calculate the concentration $C_1$ of free {{\rm P80}}
molecules in solution as $C_1=\alpha_1 C_{\rm P80}$, where $\alpha_1$
results from the numerical solution of Eq.~(2) and $C_{\rm
  P80}$ is the total concentration of {{\rm P80}}.  In particular, we
find that at the reference temperature $T_\circ$ (Fig.~S13 of the
{SM}) the {{\rm P80}} critical micellar concentration (cmc) is $0.014
\pm 0.003$~{g/L}, fully in agreement with literature~\cite{Bide2021}.

\begin{figure}
\begin{center}
\includegraphics[width=1.0\textwidth]{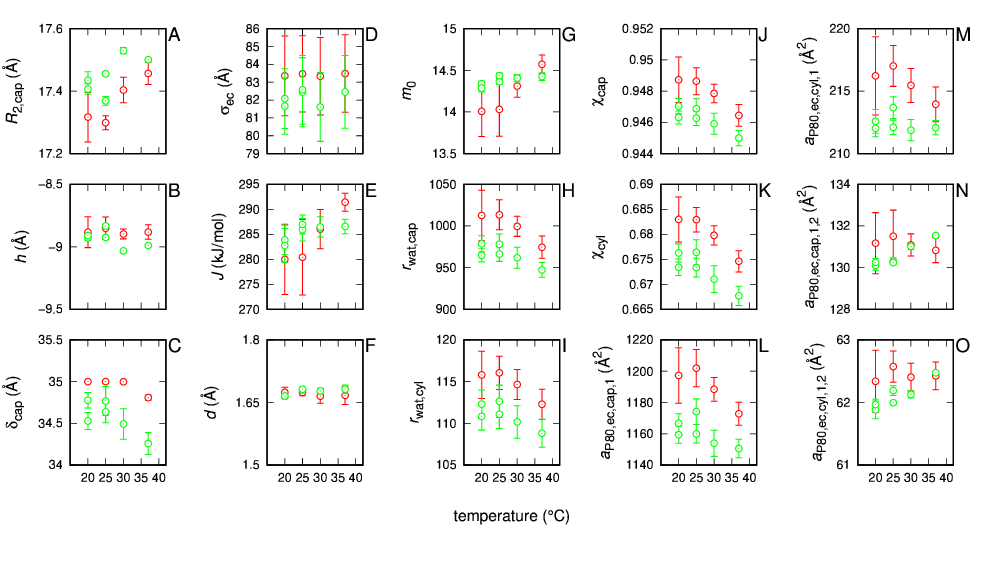}
\caption{Second-class fitting parameters (panels A-F) and derived
  fitting parameters (panels G-O) as from ESRF-SAXS data for {{\rm P80}}
    (Fig.~6A-B).
    The parameters and symbols are defined in {SM}.
    Red and green
  points refer to $C_{\rm P80}=13.3$ and 1.7~{g/L}, respectively. The validity ranges of the fit parameters shown in the panels are: A) [6,30]~{\AA};
B) [-30,30]~{\AA};
C) [6,50]~{\AA};
D) [0,100]~{\AA};
E) [0,500]~kJ/mol;
F) [0.1,10]~{\AA}.
}
  \label{diagmono}
\end{center}
\end{figure}

\subsubsection{Solid lipid nanoparticles}

We find that the SAXS curves of our LNPs
are fitted very well in the whole $q$ range by the model described in Sect.~2.2.4, paragraph ``{SAXS of stacked polydisperse platelets in the form of barrel}''  (Fig.~6C-D).
The values of the first-class
fitting parameters for the chemical groups
  have a level of uncertainty $\ll 1$~{\AA}$^3$
within reasonable validity ranges (Table~2). The mass
density of bound water, relative to bulk,  in contact with the {{\rm P80}} polar heads
 is ${\hat d}_{{\rm wat},{\rm pl}}=1.00  \pm  0.01$.
 The number of bound water molecules per
{{\rm P80}} around the platelets is $ r_{{\rm wat},{\rm P80}}\simeq  100$ (Fig.~9A), much lower than the one found in
{{\rm P80}} micelles, and the thickness of the
polar head domain is very low, $ t_1 \approx 5$~{\AA} (Fig.~9E). Therefore, SAXS data show that the bulky {{\rm P80}} polar heads
are well attached to the platelet surface. Indeed, the average
distance between two adjacent {{\rm P80}} molecules,
Eq.~S112 of the {SM}, is quite large, $d_{{\rm P80},{\rm P80}} \simeq 30$~{\AA} (Fig.~9Y) as necessary to
get a narrow coating of {{\rm P80}} polar heads.

The fitting parameters allow us to evaluate that the average platelet surface associated with each {{\rm
    P80}} molecule, Eq.~S113 of the {SM}, is $ 946 \pm 6$~{\AA}$^2$, of which $ 325 \pm 2$~{\AA}$^2$ is occupied by the {{\rm P80}} polar
head and $ 622 \pm 5$~{\AA}$^2$ by bound water.  Furthermore, according to
Eqs.~S124-S126 of the {SM},  we find that the barrel is
  consisting of $ 66.1 \pm 0.1\%$ {{\rm CP}}, $ 11.41 \pm 0.02\%$
  {{\rm P80}} and $ 22.5 \pm 0.2\%$ bound water.
In particular, we find that the small thickness of the layer
between two platelets, is almost negligible (${\Delta t}\approx
1$~{\AA}, Fig.~9M).
Considering that a water molecule's size is approximately
3~{\AA}, this finding reveals that
the
thickness ${\Delta t}\approx 1$~{\AA} must be considered as split
between two stacked P80 layers. Therefore, the bound water molecules share
this layer with P80 molecules and bridge their polar heads, as seen in
phospholipid membranes~\cite{Calero2019, Martelli2021}. Hence, two
adjacent platelets collapse one on top of the other in a stacked
conformation and comprise P80 polar heads, each hydrated by $\simeq
100$ bound water molecules. Without stacking, bound water
within the P80 layer would form an interface with unbound
water that would separate it from bulk water, as seen in phospholipid
membranes~\cite{Calero2019}. When the platelets' surface is large
enough instead, the system eliminates this interface, which would have
a free energy cost, and reduces the total free energy by stacking the
platelets into a barrel shape. Therefore, the platelets stacking is an
enthalpy-driven process with an energy-favorable mechanism provided by
the bound water.
The fitting parameters allow us to evaluate the fraction of platelets'
surface covered by {{\rm P80}} polar heads, Eq.~S114 of the {SM}, as $
\phi_{S,{\rm P80}}= 0.343 \pm 0.002$, weakly dependent of temperature
and LNP concentration. Therefore, $\approx 65$\% of the platelet
  surface is covered by  water bound to P80 and in contact with the layer of amorphous {{\rm CP}}.

\begin{figure}
\begin{center}
\includegraphics[width=0.5\textwidth]{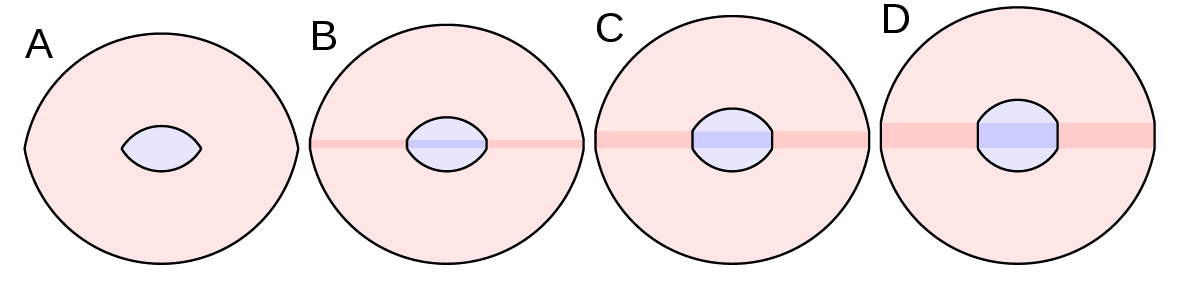}
\caption{Schematic representation of the {{\rm P80}}-micelles shapes
  as from the ESRF-SAXS data analysis. The shapes A-D are for micelles
  formed by $m=14$, 19, 24, and 29 self-assembled molecules,
  respectively. The inner azure regions represent the
    hydrophobic tails, while the pink regions represent the bulky
    hydrophilic side.}
  \label{drawmono}
\end{center}
\end{figure}

  \begin{figure}
\begin{center}
\includegraphics[width=1.0\textwidth]{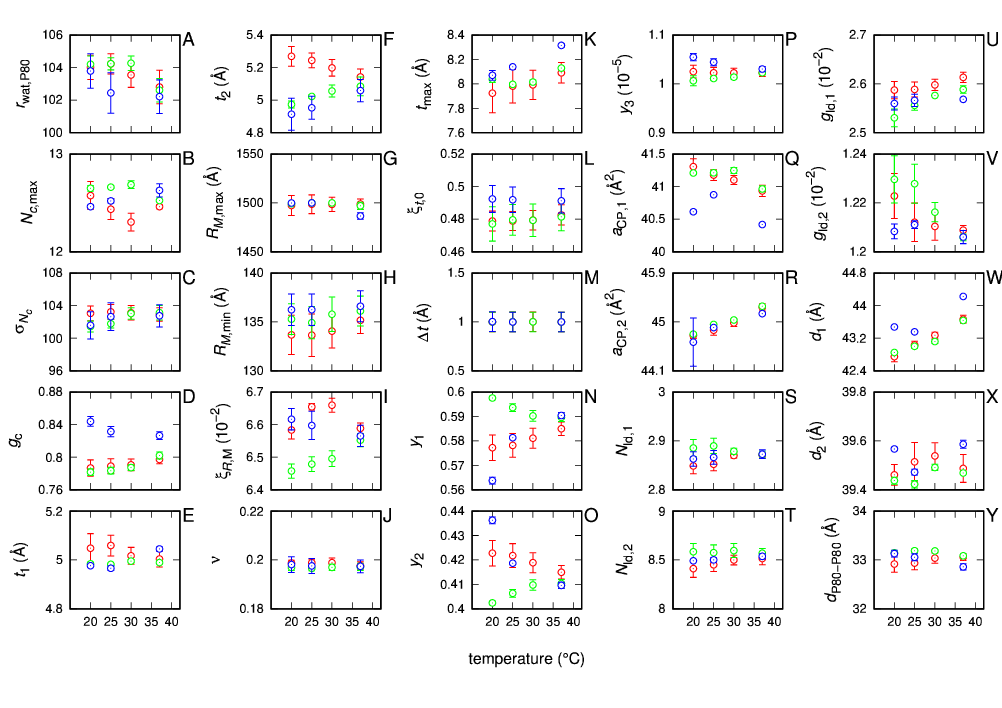}
\caption{Second-class fitting parameters (panels A, B, C, D, F, G, H,
  I, J, K, L, M, N, O, Q, R, S, T, U, V) and derived fitting
  parameters (panels E, P, W, X, Y) as from ESRF-SAXS data for
  LNPs (Fig.~6C-D).
  The parameters and symbols are defined in {SM}.
  Red, green and blue
  points refer to   LNP concentration $C_{\rm LNP}=80.0$, 40.0, and 1.0~{g/L}, respectively. The validity ranges of the fit parameters shown in the panels are: A) [35,500];
B) [10,500];
C) [2,100];
D) [0,2];
F) [4,20]~{\AA};
G) [600,3000]~{\AA};
H) [100,400]~{\AA};
I) [0,5];
J) [0,1];
K) [3,40]~{\AA};
L) [0,10];
M) [0,30]~{\AA};
N) [0,1];
O) [0,1];
Q) [30,65]~{\AA}$^2$;
R) [30,65]~{\AA}$^2$;
S) [1,20];
T) [1,20];
U) [0,1];
V) [0,1].
}
\label{diagsln}
\end{center}
\end{figure}

We estimate that the platelet has an inner radius, i.e. the radius of
the maximum circular cross-section of the barrel, with a maximum value
$R_{M,{\rm max}}\approx 1500$~{\AA} (Fig.~9G), with a
polydispersion index of $\xi_{M,{\rm max}}\approx 0.06$
(Fig.~9I), and minimum value $R_{M,{\rm min}}\approx
130$~{\AA} (Fig.~9H), with a bulging parameter $\nu
\approx 0.2$ (Fig.~9J).  Accordingly, the probability
density, $p(R)$, of the platelet radius $R$ assumes a peculiar shape
(Fig.~10A), almost independent
on temperature and LNP concentration.

The platelet core (made of {{\rm CP}}) has a maximum half-thickness
$t_{\rm max} \approx 8$~{\AA} (Fig.~9K), with a high level
of polydispersion $\xi_{t, 0}\approx 0.5$ (Fig.~9L).
Moreover, the thickness of {{\rm P80}} hydrophobic chains, embedded in
the platelet, is small $t_2\approx 5$~{\AA} (Fig.~9F).
These values allow us to calculate the probability density of the
whole platelet thickness, $2(t+t_1+t_2)$. This density
(Fig.~10D) is related to the probability density of the
half-thickness core, $p_{t}(t)$, as $p(2(t+t_1+t_2))=(1/2)p_{t}(t)$.

The distribution of the number of platelets forming a barrel-like
particle has a maximum at $N_{c, {\rm max}}\approx 13$
(Fig.~9B), with a very large standard deviation
$\sigma_{N_c}\approx 100$ (Fig.~9C) and distortion
parameter, $g_c\approx 0.8$ (Fig.~9D).
The {{\rm CP}} amorphous domain occupies a negligible part of the
platelets, $y_3\approx 10^{-5}$ (Fig.~9P), whereas the
{{\rm CP}} 1-domain accounts for almost 58\% of them, $y_1\approx
0.58$ (Fig.~9,N), with an area per molecule $a_{{\rm
    CP},1}\approx 41.5$~{\AA}$^2$ (Fig.~9Q), a repeat
distance $d_1\approx 43$~{\AA} (Fig.~9W), slightly
increasing with temperature, and a repeat number $N_{{\rm
    ld},1}\approx 3$ (Fig.~9S).
The {{\rm CP}} 2-domain occupies only $\approx 42\%$ of each platelet,
$y_2\approx 0.42$ (Fig.~9,O), with an area per molecule
$a_{{\rm CP},2}\approx 45$~{\AA}$^2$ (Fig.~9R) and a
repeat distance $d_2\approx 39.5$~{\AA} (Fig.~9X), both
increasing with temperature, and a repeat number $N_{{\rm
    ld},2}\approx 8$ (Fig.~9T).

Despite the low repeat numbers $N_{{\rm ld},1}$ and $N_{{\rm ld},2}$,
the order degree of both lamellar domains 1 and 2 is high, with
distortion parameters $g_{{\rm ld},1}\approx g_{{\rm ld},2}\approx
10^{-2}$ (Fig.~9U and~V).  The two lamellar orders agree
with similar results found by Barbosa et al.~\cite{Barbosa2018} for
LNPs composed of cetyl palmitate, Lukowski et al.~\cite{Lukowski2000},
and by Jenning and Gohla~\cite{Jenning2001}.

\begin{figure}
\begin{center}
\includegraphics[width=0.75\textwidth]{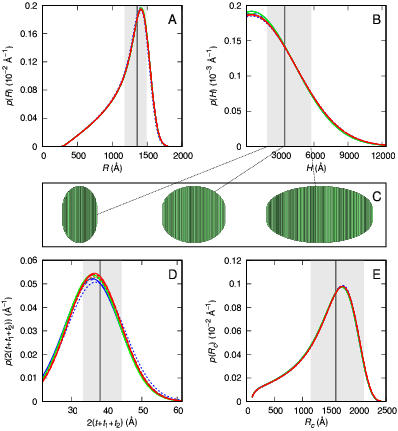}
\caption{Probability densities of the barrel circular cross-section
  radius $R$ (panel~A), the barrel height $H$ (panel~B), the total
  thickness of the platelets $2(t+t_1+t_2)$ (panel~D), and of the
  center-to-border distance $R_{\rm c}$ (panel~E), as from ESRF-SAXS
  data for LNPs.  Red, green, and blue lines refer to $C_{\rm
    LNP}=80.0$, 40.0, and 1.0~{g/L}.  Solid, dotted, and dashed lines
  refer to temperatures 20, 25, and 37~$^\circ$C.  In all panels, the
  dark-gray vertical lines indicate the median at $C_{\rm
    LNP}=80.0$~{g/L} and 20~$^\circ$C.  The shaded areas mark the
  ranges between the 1$^{\rm st}$ and the 3$^{\rm rd}$ quartile in
  each distribution.  Panel~C represents three characteristic LNPs at
  80.0~{g/L} and 20~$^\circ$C, all with $R_M$, the maximum radius of
  the circular cross-section, corresponding to the 2$^{\rm nd}$
  quartile of $p(R_M)$ distribution ($R_M=1520$~{\AA}) and with total
  height $H$ corresponding to the three quartiles of the $p(H)$
  distribution (from left to right, $1860$~{\AA}, $3390$~{\AA} and
  $5740$~{\AA}) indicated by the three dotted-lines.}
  \label{distfitsln}
\end{center}
\end{figure}

The fitting parameters also allow us to calculate the probability
densities $p(H)$ and $p(R_{\rm c})$ of the whole barrel height $H$ and
the center-to-border distance $R_{\rm c}$, respectively
(Fig.~10B and E).  We find that $H$ has a broad
distribution, with the median at 3390~{\AA}, and the 1$^{\rm st}$ and
the 3$^{\rm rd}$ quartile at 1860~{\AA} and 5740~{\AA}, respectively,
corresponding to shapes (Fig.~10C) that resemble those
observed by AFM (Fig.~5).

Nevertheless, it is essential to exercise caution when considering
$p(H)$ because data on $H\geq 10^4$~{\AA} are not directly accessible
from the experimental $q$ range. This information is derived from
various constraints, such as concentrations and molecular volumes, and
the approximations adopted in the model, e.g., the paracrystal theory.

The distribution $p(R_{\rm c})$ from our SAXS data is asymmetric and
shifted toward large values, at variance with that derived from our
AFM data (Fig.~5J) and the $p(R_H)$ from our DLS
measuraments (Fig.~4B).  The median value is at $R_{\rm c}=
1600$~{\AA}, the 1$^{\rm st}$ and the 3$^{\rm rd}$ quartiles are at
1150~{\AA} and 2100~{\AA}, respectively.  To comprehend the
inconsistencies, it's essential to consider that the three methods
have varying degrees of sensitivity regarding size. Specifically, SAXS
measurements are obtained by averaging over a significant number of
LNPs of the order of Avogadro's number, while AFM does not. As a
result, SAXS data are considered to be more dependable than
AFM. Therefore, we conclude that a representative shape for the LNPs
(Fig.~11) corresponds to the medians of the
distributions for $H$, $R_M$, and platelet thickness derived from SAXS
data (Fig.~10).

\begin{figure}
\centering\includegraphics[width=0.6\textwidth]{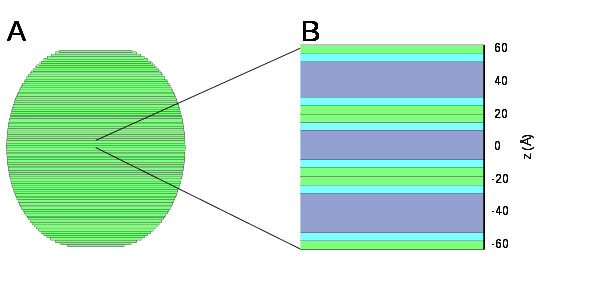}
\caption{Panel~A: Representation of a LNP with external and
    internal dimensions corresponding to the medians of the
    distributions derived from the analysis of SAXS data at $C_{\rm
      LNP}=80.0$~{g/L} and 20~$^\circ$C. (Fig.~10,
    $R_M=1520$~{\AA}, $H=3390$~{\AA},
    $2(t_0+t_1+t_2+\Delta t/2)=39$~{\AA}).
    Panel B:
    three platelets in the core, with layers schematically
    representing hydrated P80 polar heads
    (green), mixed P80 hydrophobic chains (cyan) embedded in
    amorphous CP (cyan) and lamellar CP
      (blue)). The thicknesses of the blue layers were
    sampled from the derived $p_{t}(t)$ distribution.  }
  \label{platelet_exp_2}
\end{figure}

From the SAXS measurements, we calculate, using Eq.~S39 of the {SM},
the LNP's excess electron-density ({ED}) profile along the direction
$z$ (LNP's main axis) perpendicular to three subsequent
  platelets with split distance ${\Delta t}$ (Fig.~12A).
We set the half-thickness of the {{\rm CP}} domain to its average
value $t_0$, Eq.~S31 of the {SM}. To note, the volume
  distribution functions of the hydrated P80 polar heads of a
  platelet, shown in green in Fig.~12B, are merged with that of the
  two adjacent platelets, indicating that the bound water acts as glue between the P80 polar heads belonging to two subsequent platelets. Sharp transitions in the profile
between the {{\rm P80}} polar-head domain
and the mixed {{\rm P80}} hydrophobic domain embedded in
{{\rm CP}} (shown in cyan) mark the thicknesses $t_1$ and $t_2$,
respectively. Finally, a smoother transition indicates the interface
between the latter and the {{\rm CP}} domain (shown in blue).

\begin{figure}
\begin{center}
\includegraphics[width=0.4\textwidth]{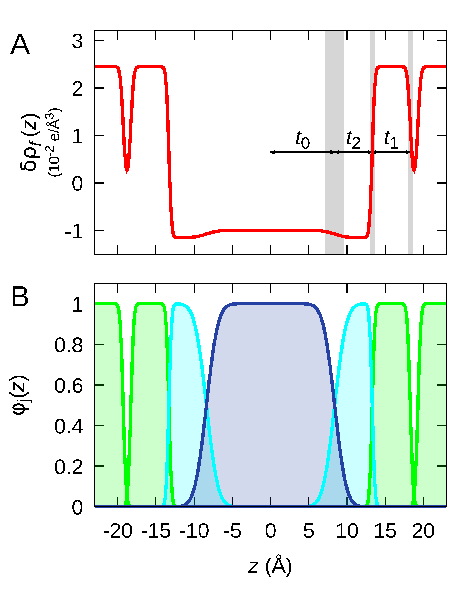}
\caption{Panel~A: Excess {ED} profile of three subsequent platelets with split-thickness ${\Delta t}$ within a LNP
  calculated from the global fit of the ESRF-SAXS data at
  $T=20^\circ$~C and $C_{\rm LNP}=80$~{g/L} (bottom curve in
  Fig.~6C-D).  From right to left, the size of the shaded
  gray bands represents the standard deviation $\sigma_{{\rm pl},j}$,
  with the indexes $j = 1$, 2, 3 corresponding to outer, middle and
  inner domain, respectively, within the region $f=3$ in Fig. 3B,.  Panel B: Volume
  fraction distributions $\phi_j(z)$, calculated according to
  Eq.~S41-S43 of the {SM}, for the hydrated {{\rm P80}} polar head
  domain (green), the mixed {{\rm P80}} hydrophobic chain domain
  embedded in  amorphous  {{\rm CP}} (cyan), and the  crystalline  {{\rm CP}} domain ({blue}).}
  \label{erf1_exp}
\end{center}
\end{figure}

\section{Conclusions}
Through synchrotron light small-angle X-ray scattering measurements at
varying temperatures and concentrations, we studied the solid
LNPs formed by {{\rm CP}} and stabilized by
{{\rm P80}}. To analyze our SAXS data, we created a novel structural
model based on data gathered from dynamic light scattering (DLS) and
atomic force microscopy (AFM) measurements. Our model effectively fit
all SAXS curves in the full scattering vector range.

Based on our findings, the shape of our LNPs is polydisperse and
barrel-like (Fig.~3A). This shape is achieved by stacking
platelets (Fig.~3B). Each platelet contains a core with
small crystalline domains of {{\rm CP}} molecules. These molecules are
elongated on the platelet surface and randomly rotated around the
normal to the surface
(Fig.~3C). The thickness of the
{{\rm CP}} core is also polydisperse, with an average thickness of
around 8~{\AA}.

Our study indicates that there are two different lamellar crystal
structures in roughly equal proportions. These structures have
characteristic (repetition) distances of approximately 43 and
39.5~{\AA}, respectively, with repeat numbers of around 3 and 8
(Fig.~3D and E).

In contrast to the standard core-shell model, we discovered that the
{{\rm P80}} molecules surround each platelet (Fig.~3B) and
intercalate between them. Their polar heads are separated by an
average distance of around 30~{\AA}, occupying a layer of
approximately 5~{\AA} with roughly 100 water molecules
bound to each head and bridging between them, with
  $\approx 1$~{\AA} between two stacked platelets.
Instead, the {{\rm P80}} apolar tails are
embedded within the amorphous CP
    portion, creating a layer roughly
5~{\AA} thick.

According to our estimations, around 35\% of the LNP's external
surface and the surface of the internal platelets are hydrophilic,
made up of {{\rm P80}} polar heads, with approximately 65\% of bound water, which favors the platelets
  stacking.
As a result, about 65\% of the barrels' volume fraction is occupied by
{{\rm CP}}, 11\% by {{\rm P80}}, and the remaining is bound water.

We believe that these findings, based on our SAXS data and the new
structural model for solid LNPs, are of
paramount importance for creating effective devices for distributing
therapeutic agents to their intended targets with improved accuracy
and precision.

  \section*{Declaration of Competing Interest}
  The authors declare that they have no known competing financial
  interests or personal relationships that could have appeared to
  influence the work reported in this paper.

\section*{CRediT authorship contribution statement}
  {\bf{Francesco Spinozzi}}: Conceptualization, Investigation, Formal analysis, Methodology, Software, Writing - Review \& Editing, Supervision.
  {\bf{Paolo Moretti}}: Investigation, Writing - Original Draft.
  {\bf{Diego Romano Perinelli}}: Investigation.
  {\bf{Giacomo Corucci}}: Investigation.
  {\bf{Paolo Piergiovanni}}: Investigation.
  {\bf{Heinz Amenitsch}}: Investigation.
  {\bf{Giulio Alfredo Sancini}}:  Conceptualization.
  {\bf{Giancarlo Franzese}}:  Conceptualization, Writing - Review \& Editing.
  {\bf{Paolo Blasi}}:  Conceptualization, Writing- Review \& Editing, Supervision, Funding acquisition.

\section*{Acknowledgments}
      The project work was funded by the ``Ministero dell'Universit\`a
      e della Ricerca'' PRIN 2017, project n. 20175XBSX4 (Targeting
      Hedgehog pathway: Virtual screening identification and
      sustainable synthesis of novel Smo and Gli inhibitors and their
      pharmacological drug delivery strategies for improved
      therapeutic effects in tumors) and partially funded by the
      European Union - Next Generation EU, Project Code: ECS00000041,
      Project Title: Innovation, digitalization and sustainability for
      the diffused economy in Central Italy - VITALITY.
      G.F. acknowledges the support by MCIN/AEI/ 10.13039/
      501100011033 and ``ERDF A way of making Europe'' grant number
      PGC2018-099277-B-C22 and PID2021-124297NB-C31.  We acknowledge
      the European Synchrotron Radiation Facility (ESRF) for provision
      of synchrotron radiation facilities and we would like to thank
      Lewis Sharpnack for assistance and support in using beamline
      ID02.  The authors thank ELETTRA for beam time allocation and
      support.

  \section*{Appendix A. Supplementary data}
  Supplementary data to this article can be found online at $\dots$
  (file \verb+SI.pdf+).

\clearpage
\bibliography{biblio_clean_clean}

\clearpage
\section*{Graphical abstract}

\begin{figure}[h]
\begin{center}
\includegraphics[width=13 cm]{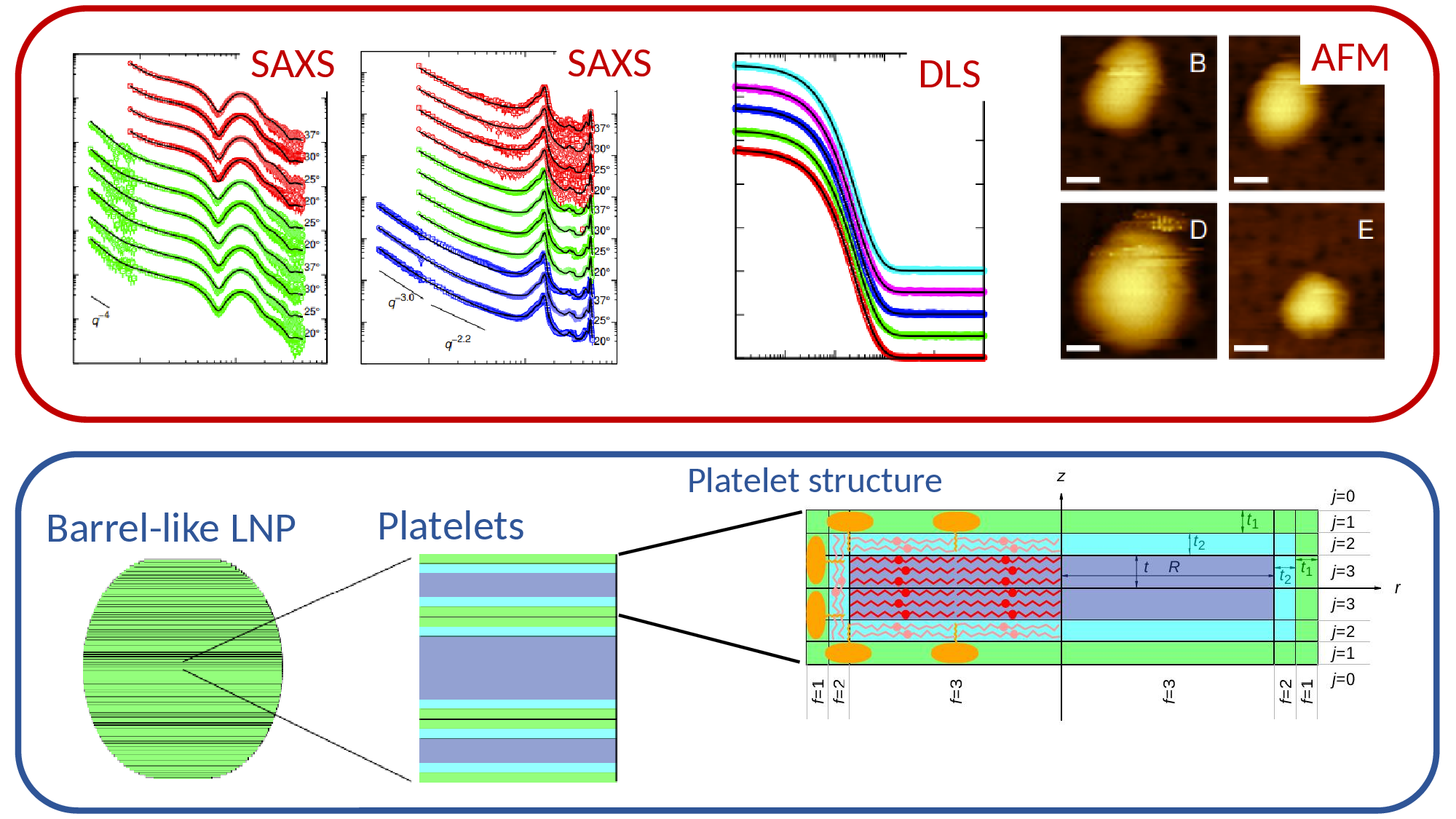}
\end{center}
\label{GA}
\end{figure}
\clearpage

\end{document}